\PassOptionsToPackage{svgnames}{xcolor}
\documentclass[twocolumn]{aastex62}

\usepackage{fixfoot}
\usepackage[clockwise, figuresright]{rotating}
\usepackage{amsmath}
\usepackage{color,soul}
\usepackage{multirow}
\usepackage{IEEEtrantools}
\usepackage{bm}
\usepackage{textcomp}
\usepackage{xcolor}
\usepackage{tikz}

\definecolor{darkgreen}{rgb}{0,0.5,0}

\received{2019 Janurary 19}
\revised{2019 April 4}
\accepted{2019 April 16}
\published{2019 May 24, \href{https://iopscience.iop.org/article/10.3847/1538-4357/ab1aa4}{\textcolor{magenta}{\apj}}, \href{https://ui.adsabs.harvard.edu/abs/2019arXiv190106509D/abstract}{877, 64}}

\shorttitle{The $(v_{\rm max},\,M_{\rm DM})$--$(\sigma_0,\,M_{\rm BH},\,\phi)$ Relations}
\shortauthors{Davis, Graham, and Combes}

\begin{document}

\title{A Consistent Set of Empirical Scaling Relations for Spiral Galaxies: The $(v_{\rm max},\,M_{\rm DM})$--$(\sigma_0,\,M_{\rm BH},\,\phi)$ Relations}

\correspondingauthor{Benjamin L. Davis}
\email{benjamindavis@swin.edu.au}

\author[0000-0002-4306-5950]{Benjamin L. Davis}
\affil{Centre for Astrophysics and Supercomputing, Swinburne University of Technology, Hawthorn, Victoria 3122, Australia}

\author[0000-0002-6496-9414]{Alister W. Graham}
\affil{Centre for Astrophysics and Supercomputing, Swinburne University of Technology, Hawthorn, Victoria 3122, Australia}

\author[0000-0003-2658-7893]{Fran\c{c}oise Combes}
\affil{Observatoire de Paris, LERMA, CNRS, PSL Universit\'e, Sorbonne Universit\'e, F-75014 Paris, France}
\affil{Coll\`ege de France, 11 Place Marcelin Berthelot, F-75005 Paris, France}


\begin{abstract}

Using the latest sample of 48 spiral galaxies having a directly measured supermassive black hole mass, $M_\text{BH}$, we determine how the maximum disk rotational velocity, $v_\text{max}$ (and the implied dark matter halo mass, $M_\text{DM}$), correlates with the (i) black hole mass, (ii) central velocity dispersion, $\sigma_0$, and (iii) spiral-arm pitch angle, $\phi$. We find that $M_\text{BH}\propto v_\text{max}^{10.62\pm1.37}\propto M_\text{DM}^{4.35\pm0.66}$, significantly steeper than previously reported, and with a total root mean square scatter ($0.58$\,dex) similar to that about the $M_\text{BH}$--$\sigma_0$ relation for spiral galaxies---in stark disagreement with claims that $M_\text{BH}$ does not correlate with disks. Moreover, this $M_\text{BH}$--$v_\text{max}$ relation is consistent with the unification of the Tully--Fisher relation (involving the total stellar mass, $M_{\rm *,tot}$) and the steep $M_{\rm BH}\propto M_{\rm *,tot}^{3.05\pm0.53}$ relation observed in spiral galaxies. We also find that $\sigma_0\propto v_\text{max}^{1.55\pm0.25}\propto M_\text{DM}^{0.63\pm0.11}$, consistent with past studies connecting stellar bulges (with $\sigma_0\gtrsim100\,\text{km\,s}^{-1}$), dark matter halos, and a nonconstant $v_\text{max}/\sigma_0$ ratio. Finally, we report that $\tan|\phi|\propto(-1.18\pm0.19)\log{v_\text{max}}\propto(-0.48\pm0.09)\log M_\text{DM}$, providing a novel formulation between the geometry (i.e., the logarithmic spiral-arm pitch angle) and kinematics of spiral galaxy disks. While the $v_\text{max}$--$\phi$ relation may facilitate distance estimations to face-on spiral galaxies through the Tully--Fisher relation and using $\phi$ as a proxy for $v_\text{max}$, the $M_\text{DM}$--$\phi$ relation provides a path for determining dark matter halo masses from imaging data alone. Furthermore, based on a spiral galaxy sample size that is double the size used previously, the self-consistent relations presented here provide dramatically revised constraints for theory and simulations.


\end{abstract}   

\section{Introduction}\label{sec:intro}

Building on the possibility of unseen mass in the solar neighborhood \citep{Jeans:1922,Kapteyn:1922,Lindblad:1926,Oort:1932}, dark matter has been considered by many astronomers to be prevalent in galaxy clusters since the 1930s \citep{Zwicky:1933,Smith:1936,Zwicky:1937,Schwarzschild:1954,Rood:1965}. In addition, the study of galaxy rotation curves provided strong evidence for the existence of dark matter \citep{Babcock:1939,Oort:1940,Freeman:1970,Rubin:1970,Rogstad:1972,Roberts:1973,Rubin:1977,Rubin:1978,Krumm:1979,Rubin:1980,Bosma:1981,Persic:1988,Broeils:1992}. The notion of nonbaryonic dark matter subsequently grew as a mechanism to explain this anomalous gravitational phenomenon \citep{Gershtein:1966,Marx:1972,Cowsik:1972,Szalay:1976}. Intriguingly, dark matter is believed to account for $84\%\pm1\%$ of the total mass in the universe \citep{Planck:2018}, but it remains elusive despite considerable efforts to achieve direct instrumental detection of its theorized particles \citep[e.g.,][]{Panda:2016,LUX:2017,XENON:2018}.\footnote{See \citet{Seigar:2015}, \citet{Arcadi:2018}, \citet{Hooper:2018}, and \citet{Salucci:2019} for reviews on searches for dark matter.} The concept of supermassive black holes (SMBHs) lurking in the center of galaxies has also had an interesting history of study \citep[see][for reviews on SMBHs]{Kormendy:1995,Longair:1996,Longair:2006,Ferrarese:Ford:2005}. It was in a spiral galaxy, our Milky Way, where an SMBH (Sagittarius A*) was first calculated to exist \citep{Lynden-Bell:1969,Sanders:1972}. For half a century, confidence in its existence gradually increased from suggestion to certainty (see \citealt{Alexander:2005} and \citealt{Genzel:2010} for reviews on Sagittarius A*), with modern measurements \citep{Gravity:2018,Gravity:2018b,Gravity:2019} even detecting the effects of general relativity \citep{Einstein:1916,Schwarzschild:1916,Kerr:1963,Bardeen:1972} in its vicinity, and providing direct imaging via very long baseline interferometry \citep{Issaoun:2019}.\footnote{See also the revolutionary imaging of the central SMBH in M87 \citep{EHT:2019}.} In contrast to dark matter, SMBHs residing at the centers of galaxies are thought to constitute a tiny fraction of the universe's total mass \citep{Graham:2007b,Vika:2009,Davis:2014,Mutlu-Pakdil:2016}.

Astronomers have long been comfortable with the idea that SMBH masses ($M_{\rm BH}$) should correlate with properties of their host galactic bulges, as evidenced by a vast literature dedicated to the study of correlations with a bulge's stellar velocity dispersion ($\sigma_0$), (baryonic and total) mass, stellar luminosity, S\'ersic index, etc.\ \citep[see][for a review]{Graham:2016b}, although many theoretical models advocate that the total gravitational mass of a galaxy or its most dominant component, the mass of its dark matter halo ($M_{\rm DM}$), should dictate the formation of SMBHs \citep[e.g.,][]{Loeb:1994,Haehnelt:1998,Silk:1998,Cattaneo:1999,Haehnelt:2000,Monaco:2000,Adams:2001}. The rotation curve informs us about the total mass of a disk galaxy and its dark matter component. Therefore, the rotational velocity profile of a galaxy disk, $v_{\rm rot}(R)$, can be considered a surrogate\footnote{Globular clusters have also been shown to be apropos tracers of the dark matter halo mass; both the globular cluster system mass \citep{Spitler:2009} and the number of globular clusters \citep{Burkert:2019} of a galaxy are directly proportional to its dark matter halo mass.} for its $M_{\rm DM}$. Thus, a relation between $M_{\rm BH}$ and some measure of $v_{\rm rot}(R)$ or $M_{\rm DM}$ should be expected in observational data.

Recently, we showed in \citet{Davis:2018} that a significant correlation exists between the total stellar mass ($M_{\rm *,tot}$) of a spiral galaxy and the mass of its central black hole. When combined with the well-known \citet{Tully:Fisher:1977} relation between the total stellar luminosity/mass of a galaxy and its rotational velocity, one obtains a relation between $v_{\rm rot}$ and $M_{\rm BH}$. A contemporary study by \citet{Tiley:2019} has defined the $z\approx0$ Tully--Fisher relation between $M_{\rm *,tot}$ and $v_{\rm rot}$ for late-type galaxies\footnote{If we calculate the Tully--Fisher relation from our dataset, we find that $M_{\rm *,tot}\propto v_{\rm rot}^{3.94\pm1.01}$, which is consistent with Equation~(\ref{eqn:TF}) at the level of $0.05\,\sigma$.} to be such that
\begin{equation}
M_{\rm *,tot}\propto v_{\rm rot}^{4.0\pm0.1}.
\label{eqn:TF}
\end{equation}
In \citet{Davis:2018}, we found that
\begin{equation}
M_{\rm BH}\propto M_{\rm *,tot}^{3.05\pm0.53}
\label{eqn:paper2}
\end{equation}
for spiral galaxies. Therefore, we should expect to find
\begin{equation}
M_{\rm BH}\propto v_{\rm rot}^{12.2\pm2.1},
\label{eqn:pred}
\end{equation}
which can subsequently be converted into an $M_{\rm BH}$--$M_{\rm DM}$ relation using an expression from \citet{Katz:2018} between $v_{\rm rot}$ and $M_{\rm DM}$ .

In addition to our focus on black hole mass scaling relations for spiral galaxies in this work, we also investigate scaling relations between $v_{\rm rot}$ (and $M_{\rm DM}$) with both $\sigma_0$ and the logarithmic spiral-arm pitch angle ($\phi$). A $v_{\rm rot}$--$\sigma_0$ ratio or relation for disk galaxies was first suggested by \citet{Whitmore:1979} and \citet{Whitmore:1981}. They found $v_{\rm rot}/\sigma_0\sim1.7$ by measuring H\textsc{i} line widths for local S0 and spiral galaxies. More recently, \citet{Cresci:2009} found that gas-rich, turbulent, star-forming $z\sim2$ disk galaxies exhibit $v_{\rm rot}/\sigma_0\sim4.4$.\footnote{The dynamics of newly assembled massive objects (DYNAMO) project \citep{Green:2010,Green:2014,Bassett:2014} demonstrated that their more local ($z\sim0.1$) sample of galaxies, selected to have high star-formation rates, are analogs of turbulent $z\simeq2$ disk galaxies, with similar $v_{\rm rot}/\sigma_0$ ratios.} The $v_{\rm rot}$--$\sigma_0$ relation represents an intriguing connection between the stellar bulge of a galaxy and its dark matter halo, although for spiral galaxies with turbulent disks, $\sigma_0$ may trace the stellar disk as much as the bulge. As a result, regular study of the $v_{\rm rot}$--$\sigma_0$ relation has persisted for 40\,yr. We contribute to this legacy with our refinement of the $v_{\rm rot}$--$\sigma_0$ relation and an original quantification of the $M_{\rm DM}$--$\sigma_0$ relation based on the $v_{\rm rot}$--$M_{\rm DM}$ expression in \citet{Katz:2018}.

As for the connection with the logarithmic spiral-arm pitch angle, \citet{Kennicutt:1981} and \citet{Kennicutt:1982} presented an observational study of the shape of spiral arms and showed that the maximum rotational velocity of a galaxy disk, $v_{\rm max}$, correlated with the spiral-arm pitch angle. They found a definitive anticorrelation, such that galaxies with higher $v_{\rm max}$ have lower $\phi$, i.e., more tightly wound spiral patterns. Their tantalizing empirical result suggests that the trend of shapes across the spiral sequence of galaxies is partly kinematic in origin. We will use our measured pitch angle values from \citet{Davis:2017} to quantify this $v_{\rm max}$--$\phi$ (and $M_{\rm DM}$--$\phi$) relation, and compare the scatter with other relations involving $\phi$.

In this work, we shall endeavor to determine scaling relations from our (currently) complete sample of spiral galaxies with directly measured black hole masses. In Section~\ref{data}, we describe our compilation of $\sigma_0$, $v_{\rm max}$, $M_{\rm BH}$, and $\phi$ measurements and detail our conversion of $v_{\rm max}$ into $M_{\rm DM}$. Section~\ref{AD} provides our regression analyses and discussion for six relations: $v_{\rm rot}$--$\sigma_0$, $M_{\rm DM}$--$\sigma_0$, $M_{\rm BH}$--$v_{\rm rot}$, $M_{\rm BH}$--$M_{\rm DM}$, $v_{\rm rot}$--$\phi$, and $M_{\rm DM}$--$\phi$. Finally, Section~\ref{end} presents our interpretation of the results and elucidates their significance. All printed uncertainties are $1\,\sigma$ ($\approx68.3\%$) confidence intervals. All magnitudes are quoted in the Vega system.

\section{Data}\label{data}

In \citet{Davis:2017}, we compiled a comprehensive sample of 44 galaxies classified as spirals and having directly measured (dynamical)\footnote{We include direct, dynamical (not including upper or lower limit) measurements derived from stellar proper motions, stellar dynamics, gas dynamics, or maser emission. We do not include black hole masses estimated via reverberation mapping, which is calibrated to the $M_{\rm BH}$--$\sigma_0$ relation \citep[e.g.,][]{Onken:2004,Peterson:2004,Graham:2011}.} black hole masses (see Table~\ref{table:Sample}). In \citet{Davis:2018,Davis:2019}, we used this sample to determine how the black hole mass of a spiral galaxy scales with its total stellar and bulge stellar masses, respectively. Here, we use the same sample, plus four new galaxies (NGC~613, NGC~1365, NGC~1566, and NGC~1672) from \citet{Combes:2019}, making our total sample of spiral galaxies with directly measured black hole masses twice the size of recent studies \citep[e.g.,][]{Sabra:2015}. For this expanded sample, we tabulate the $\phi$ and $\sigma_0$ measurements \citep[predominantly from][]{Davis:2017}, plus the $v_{\rm max}$ measurements that we have assembled from the literature. We use this rotational velocity to draw a connection to the dark matter halo mass, which dictates the maximum rotational velocity at the outer radii of a galaxy. As revealed in \citet{Davis:2017}, particular care was taken to obtain the fundamental spiral-arm pitch angle rather than the harmonics, which may be a factor of two or three smaller or larger and are accidentally obtained when an insufficiently long section of spiral arms is used.

\startlongtable
\begin{deluxetable*}{l c c r@{$\pm$}l r@{$\pm$}l r@{$\pm$}l c c r@{$\pm$}l}
\tablecolumns{9}
\tablecaption{Sample of 48 Spiral Galaxies with Directly Measured Supermassive Black Hole Masses\label{table:Sample}}
\tablehead{
\colhead{Galaxy} & \colhead{Bar?} & \colhead{$\log\left(\frac{M_{\rm BH}}{M_\sun}\right)$} & \multicolumn{2}{c}{$|\phi|$} & \multicolumn{2}{c}{$\sigma_0$} & \multicolumn{2}{c}{$v_{\rm max}$} & \colhead{$v_{\rm max}$ Reference} & \colhead{$i$} & \multicolumn{2}{c}{$\log\left(\frac{M_{\rm DM}}{M_\sun}\right)$} \\
\colhead{} & \colhead{} & \colhead{} & \multicolumn{2}{c}{(deg)} & \multicolumn{2}{c}{(km\,s$^{-1}$)} & \multicolumn{2}{c}{(km\,s$^{-1}$)} & \colhead{} & \colhead{(deg)} & \multicolumn{2}{c}{} \\
\colhead{(1)} & \colhead{(2)} & \colhead{(3)} & \multicolumn{2}{c}{(4)} & \multicolumn{2}{c}{(5)} & \multicolumn{2}{c}{(6)} & \colhead{(7)} & \colhead{(8)} & \multicolumn{2}{c}{(9)}
}
\startdata
\object{Circinus} & $\checkmark$ & $6.25^{+0.10}_{-0.12}$ & $17.0$ & $3.9$ & $148$ & $18$ & $153$ & $7$ & \citet{Courtois:2009} & 66.9 & $11.66$ & $0.12$ \\
\object{Cygnus A}\tablenotemark{a} & $\checkmark$ & $9.44^{+0.11}_{-0.14}$ & $2.7$ & $0.2$ & $270$ & $90$ & \multicolumn{2}{c}{\nodata} & \nodata & \nodata & \multicolumn{2}{c}{\nodata} \\
\object{ESO 558-G009} & & $7.26^{+0.03}_{-0.04}$ & $16.5$ & $1.3$ & $170$ & $20$ & \multicolumn{2}{c}{\nodata} & \nodata & \nodata & \multicolumn{2}{c}{\nodata} \\
\object{IC 2560} & $\checkmark$ & $6.49^{+0.19}_{-0.21}$ & $22.4$ & $1.7$ & $141$ & $10$ & $196$ & $3$ & HyperLeda & 65.6 & $11.92$ & $0.11$ \\
\object[SDSS J043703.67+245606.8]{J0437+2456} & $\checkmark$ & $6.51^{+0.04}_{-0.05}$ & $16.9$ & $4.1$ & $110$ & $13$ & \multicolumn{2}{c}{\nodata} & \nodata & \nodata & \multicolumn{2}{c}{\nodata} \\
\object{Milky Way} & $\checkmark$ & $6.60\pm0.02$ & $13.1$ & $0.6$ & $105$ & $20$ & $198$ & $6$\tablenotemark{b} & \citet{Eilers:2019} & \nodata & $11.94$ & $0.12$\tablenotemark{c} \\
\object{Mrk 1029} &  & $6.33^{+0.10}_{-0.13}$ & $17.9$ & $2.1$ & $132$ & $15$ & \multicolumn{2}{c}{\nodata} & \nodata & \nodata & \multicolumn{2}{c}{\nodata} \\
\object{NGC 0224} & $\checkmark$ & $8.15^{+0.22}_{-0.11}$ & $8.5$ & $1.3$ & $154$ & $4$ & $257$ & $6$ & HyperLeda & 72.2 & $12.21$ & $0.12$ \\
\object{NGC 0253} & $\checkmark$ & $7.00\pm0.30$ & $13.8$ & $2.3$ & $96$ & $18$ & $196$ & $3$ & HyperLeda & 75.3 & $11.93$ & $0.11$ \\
\object{NGC 0613} & $\checkmark$ & $7.57\pm0.15$\tablenotemark{d} & $15.8$ & $4.3$\tablenotemark{e} & $122$ & $18$\tablenotemark{f} & $289$ & $5$ & HyperLeda & 35.7 & $12.34$ & $0.13$ \\
\object{NGC 1068} & $\checkmark$ & $6.75\pm0.08$ & $17.3$ & $1.9$ & $174$ & $9$ & $192$ & $12$ & HyperLeda & 37.2 & $11.90$ & $0.13$ \\
\object{NGC 1097} & $\checkmark$ & $8.38^{+0.03}_{-0.04}$ & $9.5$ & $1.3$ & $195$ & $5$ & $241$ & $34$ & HyperLeda & 48.4 & $12.14$ & $0.19$ \\
\object{NGC 1300} & $\checkmark$ & $7.71^{+0.19}_{-0.14}$ & $12.7$ & $2.0$ & $218$ & $29$ & $189$ & $28$ & \citet{Mathewson:1992} & 49.6 & $11.88$ & $0.19$ \\
\object{NGC 1320} & & $6.78^{+0.24}_{-0.34}$ & $19.3$ & $2.0$ & $110$ & $10$ & $183$ & $13$ & HyperLeda & 65.8 & $11.85$ & $0.13$ \\
\object{NGC 1365} & $\checkmark$ & $6.60\pm0.30$\tablenotemark{d} & $11.4$ & $0.1$\tablenotemark{e} & $141$ & $19$\tablenotemark{f} & $198$ & $3$ & HyperLeda & 62.6 & $11.94$ & $0.11$ \\
\object{NGC 1398} & $\checkmark$ & $8.03\pm0.11$ & $9.7$ & $0.7$ & $196$ & $18$ & $289$ & $7$ & HyperLeda & 47.7 & $12.33$ & $0.13$ \\
\object{NGC 1566} & $\checkmark$ & $6.83\pm0.30$\tablenotemark{d} & $17.8$ & $3.7$\tablenotemark{g} & $98$ & $7$\tablenotemark{f} & $154$ & $14$ & \citet{Mathewson:1996} & 47.9 & $11.67$ & $0.15$ \\
\object{NGC 1672} & $\checkmark$ & $7.70\pm0.10$\tablenotemark{d} & $15.4$ & $3.6$\tablenotemark{e} & $111$ & $3$\tablenotemark{h} & $213$ & $8$ & HyperLeda & 28.2 & $12.01$ & $0.12$ \\
\object{NGC 2273} & $\checkmark$ & $6.97\pm0.09$ & $15.2$ & $3.9$ & $141$ & $8$ & $211$ & $16$ & HyperLeda & 50.1 & $12.00$ & $0.13$ \\
\object{NGC 2748} & & $7.54^{+0.17}_{-0.25}$ & $6.8$ & $2.2$ & $96$ & $10$ & $188$ & $27$ & \citet{Erroz-Ferrer:2015} & 52.9 & $11.88$ & $0.19$ \\
\object{NGC 2960} & & $7.06^{+0.16}_{-0.17}$ & $14.9$ & $1.9$ & $166$ & $16$ & $257$ & $34$ & HyperLeda & 51.5 & $12.21$ & $0.18$ \\
\object{NGC 2974} & $\checkmark$ & $8.23^{+0.07}_{-0.08}$ & $10.5$ & $2.9$ & $232$ & $4$ & $284$ & $26$ & HyperLeda & 48.1 & $12.32$ & $0.16$ \\
\object{NGC 3031} & $\checkmark$ & $7.83^{+0.11}_{-0.07}$ & $13.4$ & $2.3$ & $152$ & $2$ & $237$ & $10$ & HyperLeda & 54.4 & $12.12$ & $0.13$ \\
\object{NGC 3079} & $\checkmark$ & $6.38^{+0.11}_{-0.13}$ & $20.6$ & $3.8$ & $175$ & $12$ & $216$ & $6$ & HyperLeda & 75.0 & $12.03$ & $0.12$ \\
\object{NGC 3227} & $\checkmark$ & $7.88^{+0.13}_{-0.14}$ & $7.7$ & $1.4$ & $127$ & $6$ & $240$ & $10$ & \citet{Haynes:2018} & 59.3 & $12.14$ & $0.13$ \\
\object{NGC 3368} & $\checkmark$ & $6.89^{+0.08}_{-0.10}$ & $14.0$ & $1.4$ & $119$ & $4$ & $218$ & $15$ & HyperLeda & 46.2 & $12.03$ & $0.14$ \\
\object{NGC 3393} & $\checkmark$ & $7.49^{+0.05}_{-0.16}$ & $13.1$ & $2.5$ & $197$ & $28$ & $193$ & $48$ & \citet{Courtois:2009} & 31.8 & $11.91$ & $0.28$ \\
\object{NGC 3627} & $\checkmark$ & $6.95\pm0.05$ & $18.6$ & $2.9$ & $127$ & $6$ & $188$ & $7$ & HyperLeda & 59.2 & $11.88$ & $0.12$ \\
\object{NGC 4151} & $\checkmark$ & $7.68^{+0.15}_{-0.58}$ & $11.8$ & $1.8$ & $116$ & $3$\tablenotemark{i} & $272$ & $16$ & \citet{Mundell:1999} & 46.7 & $12.27$ & $0.14$ \\
\object{NGC 4258} & $\checkmark$ & $7.60\pm0.01$ & $13.2$ & $2.5$ & $133$ & $7$ & $222$ & $8$ & HyperLeda & 63.3 & $12.05$ & $0.12$ \\
\object{NGC 4303} & $\checkmark$ & $6.58^{+0.07}_{-0.26}$ & $14.7$ & $0.9$ & $95$ & $8$ & $214$ & $7$ & HyperLeda & 32.3 & $12.02$ & $0.12$ \\
\object{NGC 4388} & $\checkmark$ & $6.90\pm0.11$ & $18.6$ & $2.6$ & $100$ & $10$ & $180$ & $5$ & HyperLeda & 71.6 & $11.84$ & $0.11$ \\
\object{NGC 4395} & $\checkmark$ & $5.64^{+0.22}_{-0.12}$ & $22.7$ & $3.6$ & $27$ & $4$ & $145$ & $11$ & \citet{Haynes:2018} & 47.7 & $11.60$ & $0.14$ \\
\object{NGC 4501} & & $7.13\pm0.08$ & $12.2$ & $3.4$ & $166$ & $7$ & $272$ & $4$ & HyperLeda & 62.9 & $12.27$ & $0.12$ \\
\object{NGC 4594} & & $8.34\pm0.10$ & $5.2$ & $0.4$ & $226$ & $3$ & $277$ & $22$\tablenotemark{j} & HyperLeda & 47.9 & $12.29$ & $0.15$ \\
\object{NGC 4699} & $\checkmark$ & $8.34^{+0.13}_{-0.15}$ & $5.1$ & $0.4$ & $192$ & $9$ & $258$ & $7$ & HyperLeda & 42.6 & $12.22$ & $0.12$ \\
\object{NGC 4736} & $\checkmark$ & $6.78^{+0.09}_{-0.11}$ & $15.0$ & $2.3$ & $107$ & $4$ & $182$ & $5$ & HyperLeda & 31.8 & $11.84$ & $0.11$ \\
\object{NGC 4826} & & $6.07^{+0.14}_{-0.16}$ & $24.3$ & $1.5$ & $97$ & $6$ & $167$ & $9$ & HyperLeda & 55.2 & $11.75$ & $0.12$ \\
\object{NGC 4945} & $\checkmark$ & $6.15\pm0.30$ & $22.2$ & $3.0$ & $118$ & $18$ & $171$ & $2$ & HyperLeda & 77.0 & $11.78$ & $0.11$ \\
\object{NGC 5055} & & $8.94^{+0.09}_{-0.11}$ & $4.1$ & $0.4$ & $101$ & $3$ & $270$ & $14$ & \citet{Flores:1993} & 52.5 & $12.26$ & $0.13$ \\
\object{NGC 5495} & $\checkmark$ & $7.04^{+0.08}_{-0.09}$ & $13.3$ & $1.4$ & $166$ & $19$ & $202$ & $43$ & HyperLeda & 32.8 & $11.96$ & $0.25$ \\
\object{NGC 5765b} & $\checkmark$ & $7.72\pm0.05$ & $13.5$ & $3.9$ & $162$ & $19$ & $238$ & $15$ & HyperLeda & 49.1 & $12.13$ & $0.13$ \\
\object{NGC 6264} & $\checkmark$ & $7.51\pm0.06$ & $7.5$ & $2.7$ & $158$ & $15$ & \multicolumn{2}{c}{\nodata} & \nodata & \nodata & \multicolumn{2}{c}{\nodata} \\
\object{NGC 6323} & $\checkmark$ & $7.02^{+0.13}_{-0.14}$ & $11.2$ & $1.3$ & $158$ & $25$ & \multicolumn{2}{c}{\nodata} & \nodata & \nodata & \multicolumn{2}{c}{\nodata} \\
\object{NGC 6926} & $\checkmark$ & $7.74^{+0.26}_{-0.74}$ & $9.1$ & $0.7$ & \multicolumn{2}{c}{\nodata} & $246$ & $10$ & HyperLeda & 78.1 & $12.16$ & $0.13$ \\
\object{NGC 7582} & $\checkmark$ & $7.67^{+0.09}_{-0.08}$ & $10.9$ & $1.6$ & $147$ & $19$ & $200$ & $9$ & HyperLeda & 64.3 & $11.95$ & $0.12$ \\
\object{UGC 3789} & $\checkmark$ & $7.06\pm0.05$ & $10.4$ & $1.9$ & $107$ & $12$ & $210$ & $14$ & HyperLeda & 43.2 & $11.99$ & $0.13$ \\
\object{UGC 6093} & $\checkmark$ & $7.41^{+0.04}_{-0.03}$ & $10.2$ & $0.9$ & $155$ & $18$ & $170$ & $59$ & \citet{Haynes:2018} & 23.2 & $11.77$ & $0.38$ \\
\enddata
\tablecomments{
\textbf{Column~(1)}: galaxy name.
\textbf{Column~(2)}: indicates (with a checkmark) whether the galaxy exhibits a barred morphology.
\textbf{Column~(3)}: black hole mass listed in \citet{Davis:2017,Davis:2019}, compiled from references therein.
\textbf{Column~(4)}: logarithmic spiral-arm pitch angle (\emph{face-on}, absolute value in degrees) from \citet{Davis:2017}.
\textbf{Column~(5)}: central ($\lesssim0.6$\,kpc) stellar velocity dispersion listed in \citet{Davis:2017}, compiled from references therein (HyperLeda values have been updated as of December~2018).
\textbf{Column~(6)}: maximum rotational velocity of the galaxy disk (Equation~(\ref{v_max})).
\textbf{Column~(7)}: reference for maximum rotational velocity.
\textbf{Column~(8)}: inclination angle (from the reference in Column~(7), if available) used to correct the observed rotational velocity via Equation~(\ref{v_max}).
\textbf{Column~(9)}: dark matter halo mass from Equation~(\ref{eqn:conversion}); their errors should be considered minimum estimates due to lack of an intrinsic scatter measurement from \citet{Katz:2018} for Equation~(\ref{eqn:conversion}).
}
\tablenotetext{a}{Cygnus~A has been questionably classified as a spiral galaxy; it resembles a dusty early-type galaxy, although with nuclear spiral arms.}
\tablenotetext{b}{The outermost circular velocity (at a Galactocentric radius of $24.82$\,kpc) from \citet{Eilers:2019}.}
\tablenotetext{c}{Compare with the recent measurement of $\log(M_{\rm DM}/M_\sun)=11.86\pm0.01$ from \citet{Eilers:2019} for the Galaxy.}
\tablenotetext{d}{Black hole mass derived from dynamical gas measurements by \citet{Combes:2019}, who report on Atacama Large Millimeter/submillimeter Array observations of molecular tori around active galactic nuclei.}
\tablenotetext{e}{New pitch angle measurement using \textsc{2dfft} \citep{Davis:2012,2DFFT}, \textsc{spirality} \citep{Shields:2015,Spirality}, and/or \textsc{sparcfire} \citep{Davis:Hayes:2014} software packages.}
\tablenotetext{f}{From the HyperLeda database \citep{Paturel:2003}, accessed 2018~December.}
\tablenotetext{g}{From \citet{Davis:2014}.}
\tablenotetext{h}{From \citet{Garcia-Rissmann:2005}.}
\tablenotetext{i}{From \citet{Onken:2014}.}
\tablenotetext{j}{Apparent maximum stellar rotation velocity of the disk-like component of this dual morphology galaxy \citep[see][]{Gadotti:2012}.}
\end{deluxetable*}

Forty-two galaxies out of our total sample of 48 galaxies have $v_{\rm max}$ measurements in the literature. These are mostly (29 out of 42) from the HyperLeda database\footnote{\url{http://leda.univ-lyon1.fr/}} \citep{Paturel:2003}, which provides homogenized maximum rotational velocities calculated from the 21\,cm line maximum widths ($W_{\rm max}$) or available rotation curves (generally H$\alpha$ rotation curves). These velocities are derived from the observed line-of-sight velocities ($v_{\rm obs}$), after correcting for inclination ($i$), such that
\begin{equation}
v_{\rm max} = \frac{W_{\rm max}}{2\sin(i)} = \frac{v_{\rm obs}}{\sin(i)}.
\label{v_max}
\end{equation}
The internal orbital velocity for a galaxy at a given radius can be used to estimate the mass of the galaxy interior to that radius; this mass is the sum of the baryonic and nonbaryonic (i.e., dark matter) masses. At the outer regions of a galaxy, the gravitational potential is thought to be dominated by the dark matter halo (but see \citealt{Milgrom:1983} and \citealt{McGaugh:2015}), for which the maximum rotational velocity is a proxy.

For our conversions from $v_{\rm max}$ into $M_{\rm DM}$, we consulted the recent work of \citet{Katz:2018}, who determined $M_{\rm DM}$--$v_{\rm rot}$ relations from the \textit{Spitzer} Photometry and Accurate Rotation Curves (SPARC) sample gas and stellar mass models \citep{Lelli:2016} and employed the Markov chain Monte Carlo simulations from \citet{Katz:2017} to empirically determine halo masses. \cite{Katz:2018} present evidence that a variety of rotational measurements are capable of accurately predicting dark matter halo masses in spiral galaxies. They explored many of the standard measurements: (i) $v_{\rm flat}$, the rotational velocity along the flat portion of a rotation curve, (ii) $v_{2.2}$, the rotational velocity at 2.2 disk scale lengths, (iii) $v_{\rm eff}$, the rotational velocity at the effective half-light radius of a galaxy, and (iv) $v_{\rm max}$. \citet{Katz:2018} presented empirical relations for the $M_{\rm DM}$--$v_{\rm flat}$, $M_{\rm DM}$--$v_{2.2}$, $M_{\rm DM}$--$v_{\rm eff}$, and $M_{\rm DM}$--$v_{\rm max}$ relations from a sample of 120 late-type galaxies with rotational velocity measurements and dark matter halo masses estimated using the \citet{DC14} halo profile. Their halo profile model is derived from cosmological galaxy formation simulations, which account for baryonic processes affecting their host dark matter halos \citep{DC14,DC14b,Artale:2019}, unlike cosmological dark matter-only simulations \citep{NFW}.\footnote{\citet{Katz:2018} also compared their results with halo masses derived from the \citet{NFW} halo profile, but found poor results owing to ``the cusp-core problem'' for galaxies with slowly rising rotation curves \citep{Blok:2001,Blok:2002,Gentile:2004,Naray:2006,Naray:2008,Naray:2009,Katz:2017}.} \citet{Katz:2018} argue that tight relations between all of these choices for rotational velocity indicate that the kinematics of a late-type galaxy, at radii much smaller than its virial radius,\footnote{\citet{Katz:2018} note that $v_{\rm flat}$ is generally measured at a radius farther out than the typical measurement radii of $v_{2.2}$, $v_{\rm eff}$, or $v_{\rm max}$, where the halo influence dominates the rotation curve. Even at the radius ($r_{\rm flat}$) where $v_{\rm flat}$ is measured, they deduce that $\sim93\%$ (on average) of a halo mass is external to $r_{\rm flat}$.} can be reliably used to estimate its halo mass.

Based on their sample of 120 late-type galaxies, they concluded that all of these different measures produce consistent results. We have elected to use $v_{\rm max}$ due to its ubiquity in published extragalactic H\textsc{i} source catalogs. Additionally, we assume that $v_{\rm max}$ is an accurate tracer of the circular velocity ($v_{\rm circ}$) for spiral galaxies at large radii (i.e., $v_{\rm circ}\equiv v_{\rm max}$). We employ the $M_{\rm DM}$--$v_{\rm max}$ relation from \citet{Katz:2018}, which is such that
\begin{IEEEeqnarray}{rCl}
\log\left(\frac{M_{\rm DM}}{M_\sun}\right) & = & (2.439\pm0.196)\log\left(\frac{v_{\rm max}}{138\,\rm km\,s^{-1}}\right) \nonumber \\
&& +\>(11.552\pm0.108),
\label{eqn:conversion}
\end{IEEEeqnarray}
with total root mean square (rms) scatter $\Delta_{\rm rms} = 0.244$\,dex\footnote{\citet{Katz:2018} find an equivalent level of scatter ($0.242$\,dex in $\log M_{\rm DM}$) for the baryonic Tully--Fisher relation \citep{Freeman:1999,Walker:1999,McGaugh:2000} derived from the same sample of galaxies.} in the $\log M_{\rm DM}$ direction.\footnote{Because they used an unnormalized value of $v_{\rm max}$ by \citet{Katz:2018}, the error on their intercept at $v_{\rm max}=0\,\rm km\,s^{-1}$ is elevated through the inflated covariance between the slope and intercept \citep{Tremaine:2002}. In Equation~(\ref{eqn:conversion}), we have estimated a normalization constant of 138\,km\,s$^{-1}$ with an associated intercept of $11.552\pm0.108$ from the upper right panel of Figure~2 in \citet{Katz:2018}.} We apply Equation~(\ref{eqn:conversion}) to the $v_{\rm max}$ values (Table~\ref{table:Sample}, Column~(6)) and list the derived dark matter halo masses in Table~\ref{table:Sample}, Column~(9).

\vspace{5mm}

\section{Analysis and Discussion}\label{AD}

\subsection{The $v_{\rm max}$--$\sigma_0$ Relation}\label{sec:v-sigma}

The correlation between $v_{\rm rot}$ and $\sigma_0$ suggests a relationship between the dark matter halo and the stellar bulge of a galaxy, at least for galaxies with bulges. \citet{Ferrarese:2002} initially presented an empirical $v_{\rm rot}$--$\sigma_0$ relation involving 38 spiral galaxies. \citet{Baes:2003} followed up this work by adding a dozen spiral galaxies and found a consistent, tight correlation. \citet{Pizzella:2005} then argued that the $v_{\rm rot}$--$\sigma_0$ relation is different for high- and low surface brightness galaxies, while \citet[][and later \citealt{Sabra:2015}]{Ho:2007} showed from a large sample of 792 galaxies with diverse morphological types that the $v_{\rm rot}$--$\sigma_0$ relation varies significantly for subsamples based on morphology. \citet{Courteau:2007} subsequently showed that massive galaxies scattered about a $v_{\rm rot}=\sqrt{2}\sigma_0$ relation \citep[see][who found $v_{\rm rot}=1.33\sigma_0$ for early-type galaxies]{Serra:2016} and pure disks obeyed $v_{\rm rot}\simeq2\sigma_0$, and they advocated that a trivariate relationship with total light concentration was needed to properly model the data.

\begin{figure}
\includegraphics[clip=true,trim= 0mm 0mm 0mm 0mm,width=\columnwidth]{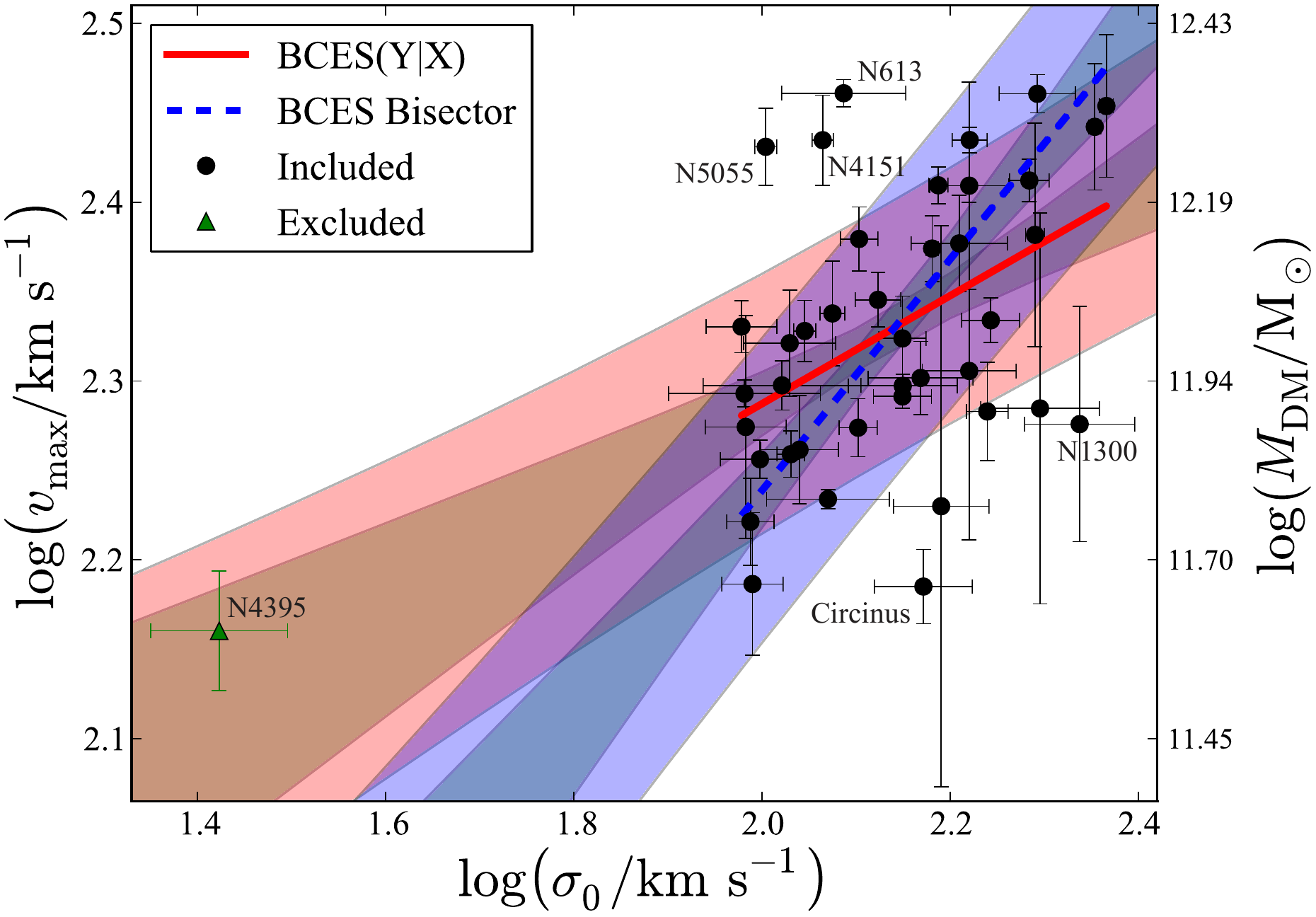}
\caption{Maximum rotational velocity (and dark matter halo mass through Equation~(\ref{eqn:conversion})) vs.\ central stellar velocity dispersion for 40 spiral galaxies (\textbullet), not including NGC~4395 (\textcolor{darkgreen}{$\blacktriangle$}). Equation~(\ref{eqn:log_v-log_sigma}) is represented by \textcolor{blue}{\tikz[baseline]{\draw[thick,dashed] (0,.5ex)--++(.5,0) ;}} and Equation~(\ref{eqn:log_v-log_sigma_alt}) is represented by \textcolor{red}{\tikz[baseline]{\draw[thick] (0,.5ex)--++(.5,0) ;}}. The dark shaded region surrounding each line shows the $\pm1\,\sigma$ uncertainty on the slope and the intercept from the regression, while the light shaded region delineates the $\pm1\,\sigma$ scatter of the data about the regression line. Error bars denote the uncertainties on $\sigma_0$ and $v_{\rm max}$, the error bars associated with $M_{\rm DM}$ are not represented and are, in actuality, larger due to error propagation. The bulgeless galaxy NGC~4395 is additionally plotted, but not included in the regressions.}
\label{fig:v-sigma}
\end{figure}

Here, we produce a $v_{\rm max}$--$\sigma_0$ relation for the 40 spiral galaxies with both $v_{\rm max}$ and $\sigma_0$ measurements (not including NGC~4395). Knowing $\sigma_0$, this relation can be used to estimate $v_{\rm max}$, and indirectly predict $M_{\rm DM}$ (through Equation~(\ref{eqn:conversion})); this is useful for galaxies with velocity dispersion measurements that lack reliable rotational velocity measurements (e.g., nearly face-on disk galaxies). The \textsc{bces} \textit{bisector} regression \citep{BCES,Nemmen:2012}\footnote{\url{https://github.com/rsnemmen/BCES}} yields
\begin{IEEEeqnarray}{rCl}
\log\left(\frac{v_{\rm max}}{\rm km\,s^{-1}}\right) & = & (0.65\pm0.10)\log\left(\frac{\sigma_0}{141\,{\rm km\,s^{-1}}}\right) \nonumber \\ 
&& +\>(2.34\pm0.01),
\label{eqn:log_v-log_sigma}
\end{IEEEeqnarray}
with $\Delta_{\rm rms} = 0.08$\,dex and intrinsic scatter\footnote{Calculated through Equation~(1) from \citet{Graham:Driver:2007}.} $\epsilon=0.07$\,dex, both in the $\log v_{\rm max}$ direction (see Figure~\ref{fig:v-sigma}).


Except for NGC~4395, it is apparent from inspection of Figure~\ref{fig:v-sigma} that five galaxies (Circinus, NGC~613, NGC~1300, NGC~4151, and NGC~5055) are mild outliers, noticeably outside the $\pm1\,\sigma$ ($\pm1\,\Delta_{\rm rms}$) scatter band. Of these five galaxies, two galaxies (NGC~1300 and NGC~5055) are $>2\,\sigma$ outliers. \citet{Davis:2018} note that NGC~1300 is an outlier in the $M_{\rm BH}$--$M_{\rm *,tot}$ diagram for spiral galaxies, with either an overmassive black hole or a low total stellar mass, which is matched here by either a low $v_{\rm max}$ or a high $\sigma_0$. \citet{Davis:2017} pointed out that NGC~5055 is an outlier in the $M_{\rm BH}$--$\sigma_0$ diagram for spiral galaxies, with either an overmassive black hole or a low central stellar velocity dispersion.

Our $(\sigma_0,\,v_{\rm max})$ dataset can be described by a Pearson correlation coefficient $r = 0.41$, and a $p$-value probability equal to $9.35\times10^{-3}$ that the null hypothesis is true. The Spearman rank-order correlation coefficient $r_s=0.40$, with $p_s=1.07\times10^{-2}$ that the null hypothesis is true. However, the slope of the $v_{\rm max}$--$\sigma_0$ relation is inconsistent with a value of 1 at the $3.5\,\sigma$ level. That is, $v_{\rm max}/\sigma_0$ is not a constant ratio. We do, however, acknowledge that the $v_{\rm max}$--$\sigma_0$ relation may be bent or curved, but we require more data with $\sigma_0<100\,{\rm km\,s^{-1}}$ to see this. We find that the slope of our $v_{\rm max}$--$\sigma_0$ relation is shallower than but consistent with the $v_{\rm rot}$--$\sigma_0$ slope ($0.84\pm0.09$) from \citet{Ferrarese:2002} for 38 spiral galaxies; the slope ($0.74\pm0.07$) from \citet{Pizzella:2005} for 40 high surface brightness disk galaxies, eight giant low surface brightness galaxies, and 24 elliptical galaxies; and the slope ($0.90\pm0.15$) from \citet{Kormendy:Bender:2011} for 30 spiral galaxies, which are a subset of the \citet{Ferrarese:2002} sample; at the levels of $1.00\,\sigma$, $0.53\,\sigma$, and $1.00\,\sigma$, respectively.

Our sample selection of only galaxies with dynamically estimated black hole masses does appear to have resulted in a selection bias, artificially truncating data such that we have a deficit of galaxies with $\sigma_0\lesssim100\,{\rm km\,s^{-1}}$ (as seen in Figure~\ref{fig:v-sigma}). \citet[][their Figure~10]{Lynden-Bell:1988} discuss this problem and present a solution by modifying the type of linear regression. By using a regression of $\log v_{\rm max}$ on $\log\sigma_0$, i.e., the \textsc{bces}(Y$|$X) routine that minimizes the residuals in the Y-direction, we can generate a relation that is less strongly affected by this bias. Doing so, we find
\begin{IEEEeqnarray}{rCl}
\log\left(\frac{v_{\rm max}}{\rm km\,s^{-1}}\right) & = & (0.30\pm0.11)\log\left(\frac{\sigma_0}{141\,{\rm km\,s^{-1}}}\right) \nonumber \\ 
&& +\>(2.33\pm0.01),
\label{eqn:log_v-log_sigma_alt}
\end{IEEEeqnarray}
with $\Delta_{\rm rms} = 0.07$\,dex and $\epsilon=0.06$\,dex, both in the $\log v_{\rm max}$ direction (see Figure~\ref{fig:v-sigma}).

Indeed, it is prudent to show concern for possible selection bias. \citet{Batcheldor:2010} pointed out that a false $M_{\rm BH}$--$\sigma_0$ relation can be obtained from galaxies with randomly assigned black hole masses by imposing a spatial resolution cutoff to censor unresolvable spheres of influence. Thus, it is important to be cautious about unobserved data in samples of spiral galaxies with dynamical black hole mass estimates. However, it is also possible that the $v_{\rm max}$--$\sigma_0$ relation breaks down for small galaxies. For example, \citet[][see their Figure~1]{Courteau:2007} show that the $v_{\rm max}$--$\sigma_0$ relation only holds for galaxies with $\sigma_0\gtrsim80\,{\rm km\,s}^{-1}$ and $v_{\rm max}\gtrsim200\,{\rm km\,s}^{-1}$, and they attribute the breakdown for smaller galaxies to less certain rotational and dispersion estimates due to amplified gas turbulence, velocity anisotropy, and measurement errors.


\subsubsection{Pseudobulges}

Galaxies that \emph{allegedly} contain pseudobulges\footnote{See \citet{Fisher:Drory:2016} for a review, summarizing the observed properties of pseudobulges and classical bulges, and see \citet{Graham:2015} for concerns with identifying pseudobulges, splitting what may be a bulge continuum, and galaxies hosting both types of bulge \citep[see][for semianalytical models of galaxy formation yielding composite bulges]{Izquierdo-Villalba:2019}. Equally concerning, \citet{Costantin:2018} assess that bulge classification is difficult with \emph{any} observational diagnostic other than their intrinsic three-dimensional shape; classical bulges appear as thick oblate spheroids, with pseudobulges distinguishable by their relative disk-like flattening.} constitute the bulk of our sample \citep[i.e., 36 out of 42;][]{Davis:2017}. For local ($z\sim0$) galaxies, the central stellar velocity dispersion is primarily a measurement of the kinematics of the bulge; however, for those galaxies with only pseudobulges or small bulges, the kinematics of the bar and disk will contribute in a higher proportion than the weak contribution in galaxies with large classical bulges \citep{Courteau:2007}.\footnote{Similarly, the use of $\sigma_e$ (the velocity dispersion within the effective half-light radius of a galaxy) is not desirable because it is heavily biased by any bar or disk in the galaxy.} \citet{Hartmann:2014} show that the central $\sigma_0$ is not only $\sigma_0({\rm bulge})$ but rather $\sigma_0({\rm bulge}+{\rm disk})$ due to contamination by the disk, which increases with disk inclination. When a classical bulge does not dominate, it is important to take into account the disk, the effect of gas (whereas the gas is negligible in massive galaxies and ellipticals), the bar orientation,\footnote{Bars, when viewed from an end-on orientation, may exhibit a small secondary biasing effect. Sight-lines down the major axis of a bar should elevate the observed $\sigma_0$, and thus more edge-on disks would have higher $\sigma_0$ values if the bar were additionally positioned in the right way within the disk plane for this sight-line to occur.} the spectral line used for the measurement,\footnote{\citet{Harris:2012} point out that the spectral line used for the measurement of $\sigma_0$ can bias the result by as much as $8\%$.} the rotation,\footnote{Spectroscopic measurements of velocity dispersion through a fixed aperture do not yield reliable estimates for the \emph{true} $\sigma_0$ of a galaxy due to contamination from galactic rotation \citep{Taylor:2010,Bezanson:2011,Uitert:2013,Hasan:2019}. \citet{Kang:2013} address the problem of rotation contaminating the $\sigma_0$ measurement and try to remove the rotation and isolate the \emph{true} $\sigma_0$ with small apertures. They report consistent $\sigma_0$ values from the optical and near-infrared absorption lines, but find that a rotation component can cause $\sigma_0$ to vary by up to $\sim20\%$ in a single-aperture spectrum.} and the inclination. These complications are beyond the scope of this project; however, we might suspect a dependency on inclination for our sample, especially for the galaxies with the lowest observed $\sigma_0$ values, where disk star contamination (\emph{if} present) will be fractionally more noteworthy. An additional investigation of the bar strength is beyond the desired scope of this work; however, a significant follow-up study might garner further insight. For example, NGC~613 and NGC~1300 (both classical barred galaxies) are prominent outliers in Figure~\ref{fig:v-sigma}, suggesting that their bars might have significantly altered the kinematics of these galaxies.

\citet{Graham:2011} demonstrate that the scatter in the $M_{\rm BH}$--$\sigma_0$ diagram can be mitigated by fitting an offset relation for barred galaxies. For pseudobulges built from bars, the ``bulge'' is very flattened, and the velocity dispersion is anisotropic, much larger in the horizontal direction than in the vertical direction. \citet{Bellovary:2014} specifically treat this anisotropic issue, and show that increased inclination not only produces an elevated scatter at the low-mass end of the $M_{\rm BH}$--$\sigma_0$ relation, but also introduces a bias, to increase globally both the average and the median value above the \emph{true} $\sigma_0$, which tends to show an offset $M_{\rm BH}$--$\sigma_0$ relation for disky galaxies. This bias applies not just to barred galaxies but all to spiral galaxies because all spiral galaxies will likely possess a bar at some point in their life. \citet{Sheth:2008} describe the development of a bar as an indicator of maturity, although bars may be recurrent and therefore not a sign of either maturity or immaturity.

In consideration of the effects of inclination, we have investigated $\sigma_0$ vs.\ $\sin(i)$ and found no correlation. We do, however, note that some galaxies have uncharacteristically low $\sigma_0$ values. For instance, NGC~253 (the Sculptor galaxy) has $\sigma_0=(96\pm18)\,{\rm km\,s^{-1}}$, but it is a bulgy, massive, and inclined ($i=75\fdg3$) galaxy. Perhaps some explanation for this galaxy can be obtained by considering that NGC~253 is a prototypical starburst galaxy \citep{Watson:1996,Kornei:2009,Leroy:2018,Mangum:2019,Martin:2019,Takano:2019}. As a result, $\sigma_0$ may depend on the age of the stars: when there is gas, young stars can form from the gas, strongly weight the optical spectrum, and ``cool'' the stellar component.\footnote{This is contrary to a collision-less simulation, where the stellar component only heats up.} This process in the disk can contribute to the measurement of $\sigma_0$, leading to a small $\sigma_0$-drop at the center of a galaxy and could explain why $\sigma_0$ is low for NGC~253 and other star-forming galaxies. Indeed, other prominent starburst galaxies display this behavior. The $v_{\rm max}/\sigma_0$ ratios of NGC~253 and three other prominent star-forming galaxies, IC~342, NGC~2146, and NGC~6946 \citep[see][and references therein]{Gorski:2018}, are all elevated with $v_{\rm max}/\sigma_0=2.04\pm0.38$, $3.11\pm0.49$, $2.33\pm0.16$, and $5.66\pm0.98$ (all from HyperLeda), respectively. Moreover, the bar in NGC~253 is not end-on but inclined (if not completely side-on), reducing its effect on $\sigma_0$.

The widespread presence of low- to medium-luminosity, active galactic nuclei (AGNs) in spiral galaxies \citep[e.g.,][]{Beifiori:2009,Jiang:2011b,Xiao:2011} indicates that the efficient growth of SMBHs through the accretion of gas is prevalent. It is evident that secular processes in disk galaxies play a key role in the development of SMBHs \citep{Cisternas:2011,Schawinski:2011,Schawinski:2012,Salvo:2012,Kocevski:2012,Simmons:2012,Treister:2012,Debattista:2013,Zubovas:2019}. Nuclear bars have been theorized to drive gas inflows into the center, feeding black holes \citep[e.g.,][]{Shlosman:1989,Shlosman:1990,Hopkins:2010}. \citet[][see also \citealt{Valluri:2016}]{Du:2017} considered the effect of an SMBH on a nuclear bar, showing that nuclear bars are destroyed at lower SMBH masses than the large-scale bars in their simulations. Specifically, they found that embedded bars are quickly dissolved when $M_{\rm BH}\gtrsim0.2\%\,M_{\rm *,tot}$, halting the inner-bar-driven gas-feeding mechanism. For comparison, we find that for the subsample of 35 pseudobulge spiral galaxies in \citet{Davis:2018}, the median ratio is is approximately an order of magnitude lower, $M_{\rm BH}=(0.016\%\pm0.011\%)\,M_{\rm *,tot}$.

The central velocity dispersion is enhanced by the presence of a bar, ultimately leading to slightly higher values of $\sigma_0$ in barred galaxies than in unbarred galaxies. \citet{Hartmann:2014} conducted simulations, showing that bar growth heats the disk, forms the bulge, and concentrates the mass,\footnote{The bar strength is not necessarily a clear indicator of mass concentration, as a galaxy could have already passed its maturity, concentrated its mass, and its once strong bar could now appear weakened or have disappeared.} therefore the mass evolution increases the central velocity dispersion,\footnote{The simulations of \citet{Hartmann:2014} show that the formation of a bar yields an elevated $\sigma_0$ even when viewing the disk face-on. This suggests a connection with out-of-plane buckling that leads to peanut shell-shaped structures \citep{Burbidge:1959,Vaucouleurs:1974,Illingworth:1983,Jarvis:1986,Combes:1990,Bureau:1999,Lutticke:2000,Lutticke:2000b,Patsis:2002,Buta:2007,Athanassoula:2016,Ciambur:2016} that are more readily visible in edge-on spiral galaxies, but are also recognizable in more face-on systems \citep{Laurikainen:2011,Laurikainen:2013,Laurikainen:2014,Laurikainen:2015,Laurikainen:2018,Erwin:2013,Athanassoula:2015,Buta:2015,Laurikainen:2016,Laurikainen:2017,Herrera-Endoqui:2017,Salo:2017,Salo:2017b,Saha:2018}. The bar or peanut shell shape may also increase the vertical dispersion of the disk, in addition to the bar, but this is a small effect considering the great streaming in the disk. Therefore, if \citet{Hartmann:2014} see an effect for $i=0\degr$, it is more likely due to the concentration of the mass in the center and the global heating than due to the peanut shell shape. Moreover, when gas is present, there is a diminished box or peanut shell-shape effect; the presence of a large gas fraction tends to weaken or destroy the peanut shell shape.} which causes barred galaxies to move to higher velocity dispersions in the $M_{\rm BH}$--$\sigma_0$ diagram. However, their simulations did not include gas \citep[see][]{Seo:2019}, which meant that gas was not present to possibly form new stars (cooling the stellar component and decreasing $\sigma_0$) or fuel SMBHs in their galaxies (increase $M_{\rm BH}$). From complementary simulations, \citet{Brown:2013} found that the growth of an SMBH also affects the nuclear stellar kinematics of a galaxy, increasing $\sigma_0$.\footnote{\citet{Hartmann:2014} remark that the increase in $\sigma_0$ due to SMBH growth within a bar, seen by \citet{Brown:2013}, is only a small effect ($\sim7\%$) compared to that resulting from the formation and evolution of a bar.} 

\subsubsection{The $M_{\rm DM}$--$\sigma_0$ Relation}\label{sec:DM-sigma}

As mentioned in Section~\ref{sec:intro}, some theoretical models suggest that dark matter halos should be related to the mass of their SMBHs. Additionally, lambda cold dark matter ($\Lambda$CDM) disk--bulge--halo models indicate that the disks and bulges are formed from the cooling of baryons inside dark matter halos, implying that the disk and bulge of a galaxy should also be partially determined by the properties of its dark matter halo \citep[e.g.,][]{Haehnelt:1998,Bosch:2000,Zhang:2000}. As a consequence of this bulge--halo connection and the known $M_{\rm BH}$--$\sigma_0$ relation, there should be an $M_{\rm DM}$--$\sigma_0$ correlation. Combining our $v_{\rm max}$--$\sigma_0$ relation (Equation~(\ref{eqn:log_v-log_sigma})) with Equation~(\ref{eqn:conversion}) yields
\begin{IEEEeqnarray}{rCl}
\log\left(\frac{M_{\rm DM}}{M_\sun}\right) & = & (1.58\pm0.29)\log\left(\frac{\sigma_0}{141\,{\rm km\,s^{-1}}}\right) \nonumber \\ 
&& +\>(12.03\pm0.12). 
\label{eqn:M_DM-sigma}
\end{IEEEeqnarray}
Additional uncertainty is introduced through the propagation of errors\footnote{Error propagation calculations were performed with the \textsc{python} package \textsc{uncertainties} (\url{http://pythonhosted.org/uncertainties/}).} when Equations~(\ref{eqn:log_v-log_sigma}) and (\ref{eqn:conversion}) are combined, however, Equation~(\ref{eqn:M_DM-sigma}) provides a useful method of quickly estimating $M_{\rm DM}$ in spiral galaxies with bulges when only $\sigma_0$ measurements are available.

\citet{Zahid:2016} examine the $\sigma_0$--$M_{\rm *,tot}$ relation for massive ($M_{\rm *,tot}\gtrsim10^{10.3}\,M_{\sun}$) quiescent galaxies at $z<0.7$. They find that $\sigma_0\propto M_{\rm *,tot}^{0.3}$, which is the same as the scaling relation between the dark matter halo velocity dispersion and halo mass (i.e., $\sigma_{\rm DM}\propto M_{\rm DM}^{0.3}$) obtained by $N$-body simulations \citep{Evrard:2008,Posti:2014}. We can compare our result by rearranging Equation~(\ref{eqn:M_DM-sigma}) to show that $\sigma_0\propto M_{\rm DM}^{0.63\pm0.11}$ for late-type galaxies has twice the slope derived by \citet{Zahid:2016} for early-type galaxies.

To the best of our knowledge, this is the first time that the relationship between the mass of the dark matter halo and the central stellar velocity dispersion has been derived or presented for spiral galaxies, although we caution against its extrapolation to late-type ($\gtrsim$Sc) spiral galaxies with $\sigma_0\lesssim100\,{\rm km\,s^{-1}}$ that potentially are overinfluenced by the disk kinematics. Alternatively, combining Equation~(\ref{eqn:log_v-log_sigma_alt}) with Equation~(\ref{eqn:conversion}) yields
\begin{IEEEeqnarray}{rCl}
\log\left(\frac{M_{\rm DM}}{M_\sun}\right) & = & (0.74\pm0.26)\log\left(\frac{\sigma_0}{141\,{\rm km\,s^{-1}}}\right) \nonumber \\ 
&& +\>(12.03\pm0.12). 
\label{eqn:M_DM-sigma_alt}
\end{IEEEeqnarray}

\subsection{The $M_{\rm BH}$--$v_{\rm max}$ Relation}

\begin{figure}
\includegraphics[clip=true,trim= 0mm 0mm 0mm 0mm,width=\columnwidth]{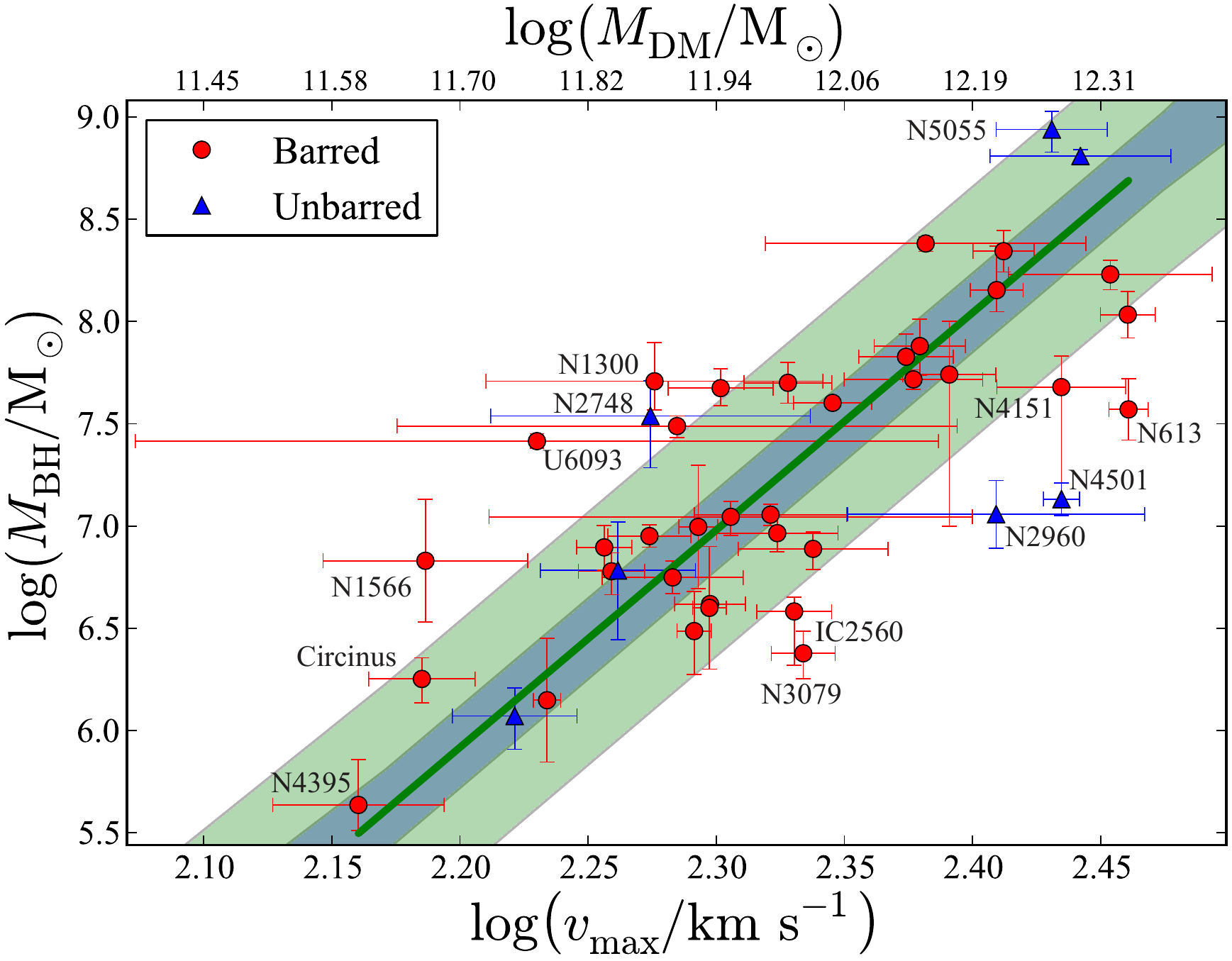}
\caption{Black hole mass vs.\ maximum rotational velocity (and dark matter halo mass according to Equation~(\ref{eqn:conversion})) for 42 galaxies. Equation~(\ref{eqn:M_BH-log_v}) is depicted by \textcolor{darkgreen}{\tikz[baseline]{\draw[thick] (0,.5ex)--++(.5,0) ;}}. Galaxies with barred morphologies are represented by \textcolor{red}{\textbullet} and those without bars are represented by \textcolor{blue}{$\blacktriangle$}.}
\label{fig:M_BH-log_v}
\end{figure}

Here, we derive the $M_{\rm BH}$--$v_{\rm max}$ scaling relation for the largest sample of spiral galaxies with directly measured black hole masses to date. Our $(v_{\rm max},\,M_{\rm BH})$ dataset, consisting of 42 galaxies, can be described by $r = 0.74$, $p=2.24\times10^{-8}$, $r_s=0.72$, and $p_s=6.70\times10^{-8}$. The \textsc{bces} \textit{bisector} regression yields the relation
\begin{IEEEeqnarray}{rCl}
\log\left(\frac{M_{\rm BH}}{M_\sun}\right) & = & (10.62\pm1.37)\log\left(\frac{v_{\rm max}}{210\,{\rm km\,s^{-1}}}\right) \nonumber \\
&& +\>(7.22\pm0.09),
\label{eqn:M_BH-log_v}
\end{IEEEeqnarray}
with $\Delta_{\rm rms} = 0.58\,{\rm dex}$ and $\epsilon=0.45\,{\rm dex}$, both in the $\log M_{\rm BH}$ direction (see Figure~\ref{fig:M_BH-log_v}). These levels of scatter place the $M_{\rm BH}$--$v_{\rm max}$ relation on par with the scatter ($\Delta_{\rm rms} = 0.63\,{\rm dex}$ and $\epsilon=0.58\,{\rm dex}$) about the $M_{\rm BH}$--$\sigma_0$ relation for spiral galaxies \citep{Davis:2017}.


The slope for the $M_{\rm BH}$--$v_{\rm max}$ relation is significantly steeper than the slopes presented in previous works. However, the steep slope is in good agreement with the prediction from Equation~(\ref{eqn:pred}), consistent at the level of $0.46\,\sigma$, which both confirms that our $M_{\rm BH}$--$v_{\rm max}$ relation is in agreement with the Tully--Fisher relation and that the $M_{\rm BH}$--$M_{\rm *,tot}$ relation must have the steep slope of $\approx3$ reported by \citet{Davis:2018}.\footnote{For reference, the $M_{\rm BH}$--(stellar bulge mass) relation for spiral galaxies has a slope of $2.44_{-0.31}^{+0.35}$ \citep{Davis:2019}.}

We do not find barred galaxies to be offset in our $M_{\rm BH}$--$v_{\rm max}$ relation (see Figure~\ref{fig:M_BH-log_v}); the bulk of our galaxies have barred morphologies (35 out of 42). \citet{Courteau:2003} found that barred galaxies are not offset in the Tully--Fisher relation, reinforcing our finding that $v_{\rm max}$ appears to predict black hole masses equally well for barred and nonbarred galaxies.\footnote{In a highly influential work, \citet{Ostriker:1973} concluded from their simulations of galaxies that massive halos would act to stabilize disks and suppress bar instabilities; thus, bars would preferentially form in galaxies with low-concentration halos. However, we now know that this is wrong. \citet{Bosma:1996} remark on the stabilizing influence of a bulge, and \citet{Sellwood:2001} find that high central rates of shear can stabilize a disk regardless of the dark matter halo density. In fact, the halo does not stabilize the disk, and even in some cases destabilizes it by increasing the bar instability \citep{Athanassoula:2002}.} That is, a single $M_{\rm *,tot}$--$v_{\rm max}$ Tully--Fisher relation for spiral galaxies combined with a single $M_{\rm BH}$--$M_{\rm *,tot}$ relation for spiral galaxies yields a single $M_{\rm BH}$--$v_{\rm max}$ relation for spiral galaxies.

Based on the behavior of four spiral galaxies, \citet{Ferrarese:2002} postulated that halos with $M_{\rm DM}\lesssim5\times10^{11}\,{M_\sun}$, i.e., $\log(M_{\rm DM}/M_\sun)\lesssim11.70$, are progressively less efficient or perhaps incapable of forming SMBHs.\footnote{Alternatively, \citet{Wassenhove:2010} find that a few percent of their simulated galaxies with massive black holes encounter stripping during their evolution, which leaves them ``naked,'' having lost most of their surrounding dark matter halos.} As can be seen in Figure~\ref{fig:M_BH-log_v}, three of our galaxies (Circinus, NGC~1566, and NGC~4395, two of which were outliers in Figure~\ref{fig:v-sigma}) have $\log(M_{\rm DM}/M_\sun) < 11.70$ (i.e., $v_{\rm max} < 159$\,km\,s$^{-1}$ from Equation~(\ref{eqn:conversion})). Our results show that $5\times10^{11}\,{M_\sun}$ is not a hard-and-fast threshold; however, given that our uncertainties on $\log M_{\rm DM}$ are greater than $\pm0.1$\,dex (see Table~\ref{table:Sample}, Column~(9)), our smallest dark matter halos are only marginally below this limit. For example, based on $v_{\rm max}=98\pm7\,{\rm km\,s^{-1}}$, we predict that NGC~1566 has a quite low $\log(M_{\rm DM}/M_\sun)=11.67\pm0.15$. However, because NGC~1566 is a Type~I Seyfert galaxy \citep{Alloin:1985,Oknyansky:2019}, it is hard to imagine it not possessing an SMBH.

If we track the location of the six outliers from Figure~\ref{fig:v-sigma} to Figure~\ref{fig:M_BH-log_v}, we find that the locations of NGC~613, NGC~1300, and NGC~4151 suggest that their maximum rotational velocities are anomalous, rather than their central stellar velocity dispersions or black hole masses. Circinus, NGC~4395, and NGC~5055 are not outliers in Figure~\ref{fig:M_BH-log_v}, indicating that they might have abnormal central stellar velocity dispersions, rather than deviant maximum rotational velocities or black hole masses. NGC~4501 is the most significant outlier in Figure~\ref{fig:M_BH-log_v}. It was not an outlier in Figure~\ref{fig:v-sigma}, implying that the black hole might be undermassive, although, it was not an outlier in any of our past black hole mass scaling relations with $\sigma_0$, $\phi$, stellar bulge mass, or stellar galaxy mass.

The $M_{\rm BH}$--$v_{\rm rot}$ relation (and/or the $M_{\rm BH}$--$M_{\rm DM}$ relation, see \S\ref{sec:BH-DM}) have been studied for nearly two decades \citep{Ferrarese:2002,Baes:2003,Zasov:2005,Sabra:2008,Kormendy:Bender:2011,Volonteri:2011,Beifiori:2012,Sun:2013,Sabra:2015}. These studies have been conducted on varied galaxy samples, including all manner of morphological types, black hole mass estimates, measures of rotational velocities, and linear regression routines. The published results for the $M_{\rm BH}$--$v_{\rm rot}$ relations have been quite diverse, ranging from a null result \citep{Kormendy:Bender:2011} to a shallow slope of $M_{\rm BH}\propto v_{\rm rot}^{2.28\pm0.67}$ \citep{Sabra:2015} and up to $M_{\rm BH}\propto v_{\rm rot}^{7.60\pm0.40}$ \citep{Volonteri:2011}. None of these results are steep enough to give consistency between the $M_{\rm BH}$--$M_{\rm *,tot}$ (Equation~(\ref{eqn:paper2})) and Tully--Fisher (Equation~(\ref{eqn:TF})) relations for spiral galaxies.

Some of the explanation for why all previous studies seem to have underestimated the slope of the $M_{\rm BH}$--$v_{\rm rot}$ relation lies in an almost ubiquitous adoption of a nearly linear $M_{\rm BH}$--$M_{\rm *,tot}$ relation. If this were true, then our earlier prediction (Equation~(\ref{eqn:pred})) would become $M_{\rm BH}\propto M_{\rm *,tot}\propto v_{\rm rot}^{4.0\pm0.1}$, which lies in the middle of the findings to date. Such a slope might conceivably make some sense for a population of massive elliptical galaxies that have formed from the result of many mergers, but even then, \citet{Sahu:2019} find $M_{\rm BH}\propto M_{\rm *,tot}^{1.65\pm0.11}$ for early-type galaxies as a whole, and, $M_{\rm BH}\propto M_{\rm *,tot}^{1.47\pm0.18}$ for core-S\'ersic early-type galaxies. A linear $M_{\rm BH}$--$M_{\rm *,tot}$ relation is therefore closer to reality for early-type galaxies, stellar orbits are more random, and both kinematical measures, $\sigma_0$ and $v_{\rm rot}$, are well correlated \citep{Zasov:2005,Courteau:2007,Ho:2007} and roughly consistent with $v_{\rm circ}=\sqrt{2}\sigma_0$ for the singular isothermal sphere profile, such that a collision-less system of stars is identical to an isothermal self-gravitating sphere of gas \citep{Binney:1987}. However, none of this applies to low-mass spiral galaxies, which have been historically underrepresented in black hole mass scaling relations.

One additional factor depressing the slopes of some past studies of the $M_{\rm BH}$--$M_{\rm *,tot}$ and $M_{\rm BH}$--$v_{\rm rot}$ relations has been the inclusion of early-type galaxies in their samples. \citet{Sahu:2019} find that early-type galaxies follow a different $M_{\rm BH}$--$M_{\rm *,tot}$ relation than late-type galaxies, with early-type galaxies with disks (ES/S0) exhibiting $M_{\rm BH}\propto M_{\rm *,tot}^{1.94\pm0.21}$. Because most early-type galaxies contain disks \citep[e.g.,][]{Emsellem:2011,Krajnovic:2011}, and lenticular galaxies also follow the Tully--Fisher relation \citep{Williams:2009}, for lenticular galaxies, Equation~(\ref{eqn:pred}) would become $M_{\rm BH}\propto v_{\rm rot}^{7.8\pm0.9}$, a result that is most similar to the results of \citet{Sabra:2008}\footnote{\citet{Sabra:2008} obtained $M_{\rm BH}\propto v_{\rm rot}^{6.75\pm0.80}$ from a sample consisting of approximately half early-type galaxies. Seven years later, \citet{Sabra:2015} found conflicting results between the relation fit from their data ($M_{\rm BH}\propto v_{\rm rot}^{2.28\pm0.67}$) and their prediction ($M_{\rm BH}\propto v_{\rm rot}^{5.15}$) from the observational study of galaxies by \citet{Bandara:2009}.} and \citet{Volonteri:2011}, consistent at the level of $0.62\,\sigma$ and $0.15\,\sigma$, respectively.

\vspace{6mm}

\subsubsection{The $M_{\rm BH}$--$M_{\rm DM}$ Relation}\label{sec:BH-DM}

Observations find AGN in galaxies with little or no bulge \citep{Satyapal:2007,Satyapal:2008} and thus hint at a possible underlying correlation between dark matter mass and black hole mass, rather than between bulge mass and black hole mass \citep{Treuthardt:2012}. Such observations elevate the possibility of an $M_{\rm BH}$--$M_{\rm DM}$ relation (with $M_{\rm DM}$ coming from $v_{\rm max}$ through Equation~(\ref{eqn:conversion})) in bulgy and bulgeless spiral galaxies, which might indicate that it is the cornerstone of black hole mass scaling relations. However, AGN feedback scenarios \citep[e.g.,][]{Debuhr:2010} suggest that the best correlation should be with the total baryonic mass rather than with the dark matter halo mass (if it is not the bulge mass) because the dark matter halo is mostly outside the baryons, and might not have much effect on the gas inflow to fuel the central SMBH. This means that the $M_{\rm BH}$--$M_{\rm DM}$ relation would be an indirect result of the $M_{\rm BH}$--(total baryonic mass) relation.

Theoretical models suggest that the $M_{\rm BH}$--$M_{\rm DM}$ relation should be nonlinear. \citet{Haehnelt:1998} demonstrated that the mass function of dark matter haloes could be matched to the $z=3$ luminosity function for quasars with lifetimes in the range of $10^6$--$10^8\,{\rm yr}$. Their models show a strong correlation between the lifetime of quasars and the slope of the $M_{\rm BH}$--$M_{\rm DM}$ relation, supporting longer quasar lifetimes as the slope of the $M_{\rm BH}$--$M_{\rm DM}$ relation becomes progressively steeper; a linear relationship would require quasar lifetimes shorter than $10^6$\,yr. \citet{Silk:1998} and \citet{Haehnelt:1998} posited that $M_{\rm BH}\propto M_{\rm DM}^{5/3}$ (here, the typical quasar lifetime is a few times $10^7$\,yr) for an isothermal sphere of cold dark matter, where a proportionality exists between the energy injected by a black hole and the gravitational binding energy ($\propto GM^2/R$) of the halo.

Alternatively, \citet{DiMatteo:2003} suggested that $M_{\rm BH}\propto M_{\rm DM}^{4/3}$, based on their $\Lambda$CDM cosmological hydrodynamic simulations. \citet{Booth:2010} concluded that $M_{\rm BH}$ is determined primarily by $M_{\rm DM}$ and found agreement with the isothermal prediction of \citet{Silk:1998} and \citet{Haehnelt:1998}, with $M_{\rm BH}\propto M_{\rm DM}^{1.55\pm0.05}$, which is itself in exact agreement with the observational study of \citet{Bandara:2009}, who find $M_{\rm BH}\propto M_{\rm DM}^{1.55\pm0.31}$.\footnote{See also the observational study by \citet{Bogdan:2015}, who find $M_{\rm BH}\propto M_{\rm DM}^{1.6_{-0.4}^{+0.6}}$ from a sample of 3130 elliptical galaxies. } However, these results should not be considered applicable to spiral galaxies as these studies focused on massive galaxies, e.g., the sample of \citet{Bandara:2009} consisted of galaxies with $M_{\rm BH}\gtrsim10^8\,M_\sun$ and $M_{\rm DM}\gtrsim10^{13}\,M_\sun$. Other published slopes for this relation, from galaxies with $M_{\rm BH}$ estimates, include $M_{\rm BH}\propto M_{\rm DM}^\text{1.65--1.82}$ \citep{Ferrarese:2002} for 38 spiral galaxies and $M_{\rm BH}\propto M_{\rm DM}^{1.27}$ \citep{Baes:2003} for 50 spiral galaxies.

Recently, \citet{Burcin:2018} found $M_{\rm BH}\propto M_{\rm DM}^{1.55\pm0.02}$ from a new set of simulations \citep{Vogelsberger:2014}, which is in precise agreement with the above findings \citep{Bandara:2009,Booth:2010}. Their work is different from previous studies because it focused on extending the relation to lower-mass spiral galaxies. However, their precise agreement with past studies of high-mass galaxies and an isothermal model indicate that their simulations remain tied or tuned to the old black hole mass scaling relations that assume that black holes and their host galaxies follow a nearly linear $M_{\rm BH}$--$M_{\rm *,tot}$ relation. Specifically, \citet{Burcin:2018} perceive that $M_{\rm BH}\propto M_{\rm *,tot}^{1.53\pm0.02}$, with a slope half as steep as our result for late-type galaxies (Equation~(\ref{eqn:paper2})), but consistent with the relation $M_{\rm BH}\propto M_{\rm *,tot}^{1.65\pm0.11}$ for early-type galaxies from \citet{Sahu:2019}.\footnote{From inspection of Figures~7 and 8 from \citet{Burcin:2018}, we point out an apparent bend in the respective scaling relations with black hole mass. There is a noticeable dichotomy between low- and high-mass black holes, such that it warrants a steeper slope for spiral galaxies hosting low-mass black holes and a shallow slope for early-type galaxies hosting high-mass black holes, rather than their single regression fit to the combined sample.} Combining the results of \citet{Burcin:2018}, $M_{\rm BH}\propto M_{\rm *,tot}^{1.53\pm0.02}\propto M_{\rm DM}^{1.55\pm0.02}$, would suggest that the $M_{\rm *,tot}$--$M_{\rm DM}$ relation is almost exactly linear.

Evolutionary studies show how the $M_{\rm *,tot}$--$M_{\rm DM}$ relation changes with time \citep{Hansen:2009,Behroozi:2010,Behroozi:2013,Behroozi:2018,Moster:2010,Moster:2013,Moster:2018,Leauthaud:2012,Yang:2012,Reddick:2013,Wang:2013,Birrer:2014,Lu:2015,Puebla:2017,Kravtsov:2018,Mowla:2019}. \citet{Tiley:2019} performed an analysis of the Tully--Fisher relation at $z\approx0$ and $z\approx1$ by carefully comparing local and distant samples of late-type galaxies and degrading the quality of the local sample to match that of the distant sample. They concluded that no significant difference is apparent between the two epochs, suggesting an intimate link between stellar mass and dark matter (as traced by $v_{\rm max}$) in late-type galaxies over at least an $\approx8$\,Gyr period.

The $M_{\rm *,tot}$--$M_{\rm DM}$ relation is traditionally fit with a double power law, exhibiting a steep low-mass slope and a shallow high-mass slope \citep{Yang:2012,Moster:2013}.\footnote{However, \citet{Behroozi:2013} note that a pure double power law is not perfect and results in a $M_{\rm BH}$--$M_{\rm DM}$ relation that is off by as much as $0.1$\,dex.} For the low-mass portion of the double power-law $M_{\rm *,tot}$--$M_{\rm DM}$ relation, \citet{Katz:2018b} speculate that late-type galaxies should follow a theoretical $M_{\rm *,tot}\propto M_{\rm DM}^{5/3}$ relation, powered by supernovae feedback and its role in regulating galaxy formation. Furthermore, \citet{Katz:2018b} follow this theoretical prediction up with an observational determination of $M_{\rm *,tot}\propto M_{\rm DM}^{1.51\pm0.13}$ from their sample of late-type galaxies from the SPARC sample with \citet{DC14} dark matter halo profiles. Combining this result with Equation~(\ref{eqn:paper2}) necessarily implies that
\begin{equation}
M_{\rm BH}\propto M_{\rm DM}^{4.61\pm0.89}.
\label{eqn:pred2}
\end{equation}
Combining our spiral galaxy $M_{\rm BH}$--$v_{\rm max}$ relation (Equation~(\ref{eqn:M_BH-log_v})) with Equation~(\ref{eqn:conversion}) yields
\begin{IEEEeqnarray}{rCl}
\log\left(\frac{M_{\rm BH}}{M_\sun}\right) & = & (4.35\pm0.66)\log\left(\frac{M_{\rm DM}}{10^{12}\,{M_\sun}}\right) \nonumber \\
&& +\>(7.22\pm0.12),
\label{eqn:M_BH-M_DM}
\end{IEEEeqnarray}
which is consistent with the slope from Equation~(\ref{eqn:pred2}) at the level of $0.17\,\sigma$. This slope for spiral galaxies is notably steeper than reported in past studies \citep[e.g.,][]{Haehnelt:1998,Silk:1998,Ferrarese:2002,Baes:2003,Bandara:2009,Booth:2010,Bogdan:2015,Burcin:2018}.

\subsection{The $v_{\rm max}$--$\phi$ Relation}

The maximum rotational velocity does not only correlate with $M_{\rm BH}$ (Figure~\ref{fig:M_BH-log_v}), but it also correlates with, among other things, the spiral-arm pitch angle, $\phi$ \citep[for an introduction, see][]{Davis:2017}.\footnote{Logarithmic spirals have been adopted as the natural metric for the form of spiral arms in disk galaxies since the pioneering works of \citet{Pahlen:1911}, \citet{Groot:1925}, \citet{Reynolds:1925}, \citet{Lindblad:1927,Lindblad:1938,Lindblad:1941}, and \citet{Danver:1942}.} \citet{Kennicutt:1981} observed a correlation between the maximum rotational velocity and the spiral-arm pitch angle, which was later reanalyzed and quantified by \citet{Savchenko:2011}. \citet{Kennicutt:1981} described the anticorrelation as being ``fairly strong,''\footnote{\citet{Kendall:2011,Kendall:2015} present a study of $\phi$ vs.\ $v_{\rm flat}$ from near-infrared \textit{Spitzer Space Telescope} imaging and ``find no evidence for the strong anticorrelation'' that was found in the optical by \citet{Kennicutt:1982}.} and claimed that it supports the density-wave theory expectations for spiral-arm geometry \citep{Roberts:1975}.\footnote{See the recent reviews on spiral structure in disk galaxies \citep{Dobbs:2014} and spiral density wave theory \citep{Shu:2016} for additional information and historical context.}

\begin{figure}
\includegraphics[clip=true,trim= 0mm 0mm 0mm 0mm,width=\columnwidth]{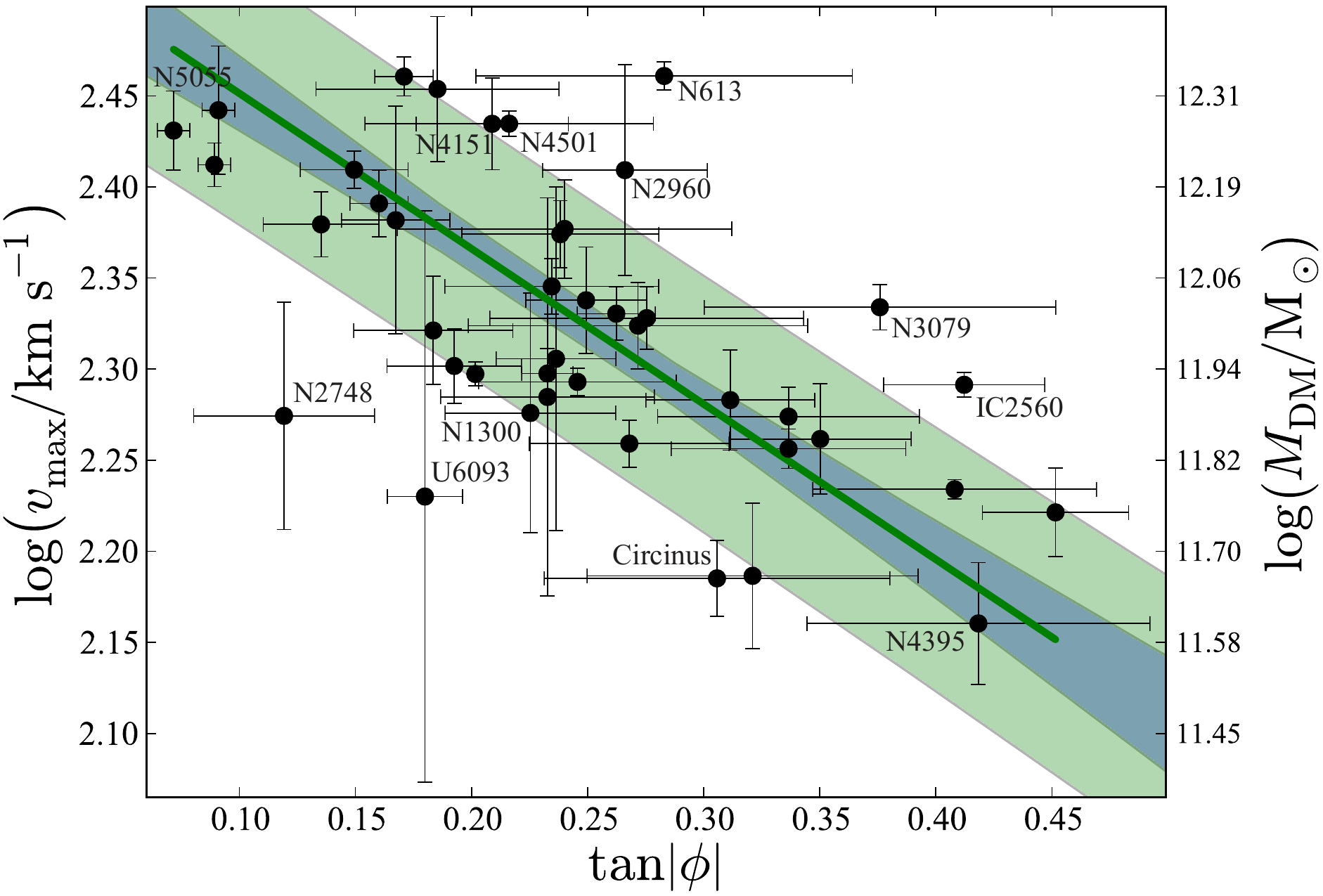}
\caption{Maximum rotational velocity (and dark matter halo mass according to Equation~(\ref{eqn:conversion})) vs.\ the tangent function of the spiral-arm pitch angle absolute value for our sample of 42 spiral galaxies. Equation~(\ref{eqn:phi-log_v_rot}) is represented by \textcolor{darkgreen}{\tikz[baseline]{\draw[thick] (0,.5ex)--++(.5,0) ;}}.}
\label{fig:phi-v_rot}
\end{figure}

Our $(\tan|\phi|,\,\log v_{\rm max})$ dataset of 42 galaxies can be described by $r = -0.62$, $p=1.31\times10^{-5}$, $r_s=-0.58$, and $p_s=5.73\times10^{-5}$. The \textsc{bces} \textit{bisector} regression yields
\begin{IEEEeqnarray}{rCl}
\log\left(\frac{v_{\rm max}}{\rm km\,s^{-1}}\right) & = & (-0.85\pm0.14)\left[\tan|\phi|-\tan13\fdg35\right] \nonumber \\
&& +\>(2.33\pm0.01),
\label{eqn:phi-log_v_rot}
\end{IEEEeqnarray}
with $\Delta_{\rm rms} = 0.07$\,dex and $\epsilon=0.05$\,dex, both in the $\log v_{\rm max}$ direction (see Figure~\ref{fig:phi-v_rot}).


The outliers in Figure~\ref{fig:phi-v_rot} match the behavior in the $M_{\rm BH}$--$v_{\rm max}$ diagram (Figure~\ref{fig:M_BH-log_v}). This reflects the tight $M_{\rm BH}$--$\phi$ relation \citep{Seigar:2008,Berrier:2013,Davis:2017}. However, NGC~2748 was an outlier in the $M_{\rm BH}$--$\phi$ diagram \citep{Davis:2017}, suggesting that its continued deviance in the $v_{\rm max}$--$\phi$ diagram (Figure~\ref{fig:phi-v_rot}) is more due to its irregular spiral arms than to its maximum rotational velocity measurement because it was not an outlier in the $v_{\rm max}$--$\sigma_0$ diagram (Figure~\ref{fig:v-sigma}).

\citet{Kennicutt:1981} qualitatively investigated the $v_{\rm max}$--$\phi$ relation using a sample of 113 spiral galaxies, but he did not perform a linear regression. \citet{Savchenko:2011} culled the sample from \citet{Kennicutt:1981} down to a subsample of 46 galaxies and performed a linear (not a power-law) regression, yielding
\begin{equation}
|\phi|=\left(-0.049\pm0.008\right)\left(\frac{v_{\rm max}}{\rm km\,s^{-1}}\right)+\left(22\fdg85\pm1\fdg76\right).
\label{eqn:Savchenko}
\end{equation}
When we perform a similar regression on our ($v_{\rm max},\,|\phi|$) dataset of 42 spiral galaxies, we find that the \textsc{bces} \textit{bisector} regression yields
\begin{IEEEeqnarray}{rCl}
|\phi| & = & \left(-0.127\pm0.019\right)\left(\frac{v_{\rm max}-210\,{\rm km\,s^{-1}}}{{\rm km\,s^{-1}}}\right) \nonumber \\
&& +\>\left(14\fdg43\pm0\fdg66\right),
\label{eqn:SavCompare}
\end{IEEEeqnarray}
with $\Delta_{\rm rms} = 4\fdg36$ and $\epsilon=3\fdg48$ in pitch angle. Our slope of $-0.127\pm0.019$ is more than $2.5$ times as steep as their slope. However, we argue that the slope of \citet{Savchenko:2011} in Equation~(\ref{eqn:Savchenko}) is too shallow because their zero-point predicts that $v_{\rm max}=0$ at $|\phi|=22\fdg85\pm1\fdg76$. In contrast, our zero-point would predict that $v_{\rm max}=0$ at $|\phi|=41\fdg11\pm3\fdg92$, which is on par with some of the highest measured pitch angle measurements in the literature. Nevertheless, such a formulation of the $v_{\rm max}$--$\phi$ relation is inherently limited, given that $v_{\rm max}$ equals zero at high $\phi$, which should not be the case. Only a rotating disk can have a spiral pattern with a measurable pitch angle. We prefer our formulation of the $v_{\rm max}$--$\phi$ relation in Equation~(\ref{eqn:phi-log_v_rot}), where $v_{\rm max}$ never equals zero. Because $\phi$ is defined such that $0\degr\leq|\phi|\leq90\degr$, the best-fit values from Equation~(\ref{eqn:phi-log_v_rot}) yield $340\geq v_{\rm max}/{\rm km\,s^{-1}}>0$, respectively.

\citet{Savchenko:2011} used the $z\sim0$ correlation to predict $v_{\rm max}$ for a distant sample of galaxies in the \textit{Hubble Deep Fields} \citep{Ferguson:2000,Beckwith:2006} and ultimately concluded that the distant galaxies follow the local Tully--Fisher relation if they follow the local $v_{\rm max}$--$\phi$ relation that they derived (for which we just doubled the slope). Given that \citet{Tiley:2019} observe no significant difference between the zero-points\footnote{\citet{Tiley:2019} remark that it is difficult to precisely constrain the slope of the Tully--Fisher relation at $z\approx1$ due to large scatter; they instead perform fixed slope fits to their $z\approx1$ sample.} of the Tully--Fisher relation at $z\approx0$ and $z\approx1$,\footnote{This lack of evolution in the zero-point is also seen in the baryonic Tully--Fisher relation out to $z=0.6$ \citep{Puech:2010,Puech:2011} and possibly even to $z=1.2$ \citep{Weiner:2006}.} we combine our modified $v_{\rm max}$--$\phi$ relation (Equation~(\ref{eqn:phi-log_v_rot})) with the $z\approx0$ $K$-band absolute magnitude ($\mathfrak{M}_K$)\footnote{Near-infrared imaging can offer a more complete view of spiral morphology as the optical morphologies of some spiral galaxies can appear significantly different \citep{Block:1991,Block:1994,Seigar:1998,Seigar:1998b}, with flocculent optical morphologies occasionally exhibiting grand-design morphologies in the near-infrared \citep{Thornley:1996,Seigar:2003}. On the whole, \citet{Buta:2010} found that (save for the most dusty galaxies) $3.6\,\micron$ morphological classifications are well correlated with blue-light classifications.} Tully--Fisher relation from \citet{Tiley:2019}
\begin{IEEEeqnarray}{rCl}
\mathfrak{M}_K & = & (-8.3\pm0.3)\log\left(\frac{v_{\rm rot}}{100\,{\rm km\,s^{-1}}}\right) \nonumber \\
&& -\>(22.26\pm0.07)\,{\rm mag}, 
\label{eqn:phi-mag}
\end{IEEEeqnarray}
with $\Delta_{\rm rms} = 1.17$\,mag and $\epsilon=1.13$\,mag, to obtain
\begin{IEEEeqnarray}{rCl}
\mathfrak{M}_K & = & (7.08\pm1.17)\left[\tan|\phi|-\tan13\fdg35\right] \nonumber \\
&& -\>(25.03\pm0.15)\,{\rm mag}. 
\label{eqn:phi-mag}
\end{IEEEeqnarray}
This new relation allows us to estimate the intrinsic luminosity of a spiral galaxy (and therefore, a distance) based on its spiral geometry \citep[see][who presented a strong correlation between absolute luminosity and the degree of development of spiral arms]{Bergh:1960a,Bergh:1960b}, which is measurable from simple imaging.

\subsubsection{The $M_{\rm DM}$--$\phi$ Relation}

Here, we present new evidence that spiral galaxy geometry becomes a useful indicator for the dark matter content of a galaxy. Combining our updated $v_{\rm max}$--$\phi$ relation (Equation~(\ref{eqn:phi-log_v_rot})) with the $M_{\rm DM}$--$v_{\rm max}$ relation (Equation~(\ref{eqn:conversion})) yields
\begin{IEEEeqnarray}{rCl}
\log\left(\frac{M_{\rm DM}}{M_\sun}\right) & = & (-2.08\pm0.37)\left[\tan|\phi|-\tan13\fdg35\right] \nonumber \\
&& +\>(12.03\pm0.11).
\label{eqn:M_DM-phi}
\end{IEEEeqnarray}
Adding the total rms scatters from Equations~(\ref{eqn:conversion}) and (\ref{eqn:phi-log_v_rot}) in quadrature yields an implied $\Delta_{\rm rms} \simeq 0.25\,{\rm dex}$ in the $\log M_{\rm DM}$ direction. This level of scatter is equivalent to that of the $M_{\rm *,tot}$--$\phi$ relation from \citet{Davis:2018}, indicating a connection between the large gravitational potential well of a galaxy and its spiral-arm geometry, which is manifest in the $v_{\rm max}$--$\phi$ relation (Equation~(\ref{eqn:phi-log_v_rot})). The ability of $\phi$ to predict dark matter halo masses in late-type galaxies with such a small slope and level of scatter is remarkable given that no other parameter in this paper is as simple to measure or has such minimal observational requirements.

\citet{Burcin:2018} also presented an $M_{\rm DM}$--$\phi$ relation by measuring the pitch angles of spiral galaxies produced from the \textit{Illustris} simulation \citep{Vogelsberger:2014}. They found from 95 simulated spiral galaxies that $M_{\rm DM}\propto(-0.029\pm0.001)|\phi|$. When we fit a linear \textsc{bces} \textit{bisector} regression between $M_{\rm DM}$ and $|\phi|$ for our sample of 42 spiral galaxies, we similarly find $M_{\rm DM}\propto(-0.028\pm0.007)|\phi|$. Therefore, because we find agreement between our observations and the \textit{Illustris} simulations for the $M_{\rm DM}$--$\phi$ relation (Equation~(\ref{eqn:M_DM-phi})), but disagree with \citet{Burcin:2018} when it comes to the $M_{\rm BH}$--$M_{\rm DM}$ relation (Equation~(\ref{eqn:M_BH-M_DM})), we conclude that the \textit{Illustris} simulation (and indeed most simulations) fail to properly model black hole mass scaling relations for spiral galaxies. Specifically, the simulations appear to correctly model the total stellar mass of spiral galaxies and the corresponding geometry of their spiral arms but underestimate the mass of SMBHs in spiral galaxies.

The existence of an $M_{\rm DM}$--$\phi$ relation (Equation~(\ref{eqn:M_DM-phi})) is not unforeseen; the $M_{\rm DM}$--$\phi$ and $M_{\rm BH}$--$\phi$ relations can be seen as complementary relations to the $M_{\rm BH}$--$M_{\rm DM}$ relation (Equation~(\ref{eqn:M_BH-M_DM})). Moreover, a correlation has been known to exist between rotation curve shear rate ($S$)\footnote{Shear is a dimensionless parameter that quantifies the shape (rising, falling, or flat) of rotation curves, and thus is dependent on the total baryonic and dark matter concentrations. $S=0.5$ for a flat rotation curve, $S>0.5$ for a falling rotation curve, and $S<0.5$ for a rising rotation curve \citep[see][for a formal definition]{Seigar:2014}.} and spiral-arm pitch angle. \citet{Block:1999} and \citet{Seigar:2004,Seigar:2005,Seigar:2006,Seigar:2014} have demonstrated the existence of an $S$--$\phi$ relation from observations, and \citet{Grand:2013} have confirmed the correlation through $N$-body simulations. Precisely, $S=(0.88\pm0.08)-(0.014\pm0.001)|\phi|$ \citep{Seigar:2006} and $S\simeq0.85-0.014|\phi|$ \citep{Grand:2013}.\footnote{See also the additional studies of the $S$--$\phi$ relation from simulations \citep{Michikoshi:2014} and observations \citep{Kendall:2015,Yu:2019}. Similarly, \citet{Font:2019} conducted a comparison of the ``shear parameter'' (the difference between the pattern speed of the bar and the spiral arms, relative to the angular rate of the outer disk) vs.\ $\phi$ in their observational study.}

\citet{Kalinova:2017} introduced a new parameter to classify the rotation curves of galaxies quantitatively: the coefficient of the first eigenvector (PC$_1$) of the reconstructed circular velocity rotation curve of a galaxy through principal component analysis.\footnote{PC$_1$ is a measure of the shape and amplitude of the rotation curve and thus reflects the mass distribution of dark matter in galaxies. Galaxies with PC$_1>0$ exhibit high-amplitude centrally-peaked rotation curves and galaxies with PC$_1<0$ exhibit low-amplitude slow-rising rotation curves \citep[see][for additional information]{Kalinova:2017}.} \citet{Yu:2019} performed an observational study of PC$_1$ vs.\ $|\phi|$ and uncovered a strong correlation ($r=-0.66$), from which they deduced the implication ``that dark matter content might help to shape spiral arms.''

While perhaps not fundamental, $\phi$ is more a consequence of the presence of a sizeable stabilizing mass, making the disk less self-gravitating. Indeed, when the disk is stable, due to a large central mass (contributed by the bulge and maybe also the dark matter halo), then the pitch angle is small with a tightly wound spiral pattern or even a multi-arm structure. When the disk is unstable, it forms a bar, which can emanate loosely wound spiral arms from its ends. The observational results from \citet{Buta:2010b} support this, showing that the frequency of strong bars progressively increases when moving through the morphological types from lenticular to early- and late-type spirals. Additionally, there can be evolution in time, since violent bar instability heats the disk, which becomes more stable due to a high Toomre parameter, $Q\equiv\sigma_v/\sigma_{\rm crit}$, where $\sigma_v$ is the tangential velocity dispersion and $\sigma_{\rm crit}$ is the critical (i.e., the minimum dispersion needed for stability) velocity dispersion. \citet{Toomre:1964} established that $\sigma_{\rm crit}\propto\mu/\kappa$, where $\mu$ is the disk surface density, and $\kappa$ is the epicyclic frequency; $\kappa$ is particularly high when there is a high mass concentration, making $\sigma_{\rm crit}$ small and the disk stable. In the end, what counts is the mass concentration, not only the bulge (which is part of it). Rotational velocity is also a sign of mass concentration; $v_{\rm max}$ is higher if the mass is more concentrated for the same total mass.

Numerical simulations also illustrate the importance of mass concentration. A sufficient mass concentration in barred galaxies creates two inner Lindblad resonances, inside of which the stable (x2) periodic prograde orbits\footnote{For supplementary reading on families of periodic orbits, see \citet{Contopoulos:1980}, \citet{Contopoulos:1989}, \citet{Sellwood:1993}, \citet{Binney:1987}, and \citet{Sellwood:2014b}.} are perpendicular to the bar and thus do not sustain the bar, but rather weaken it \citep{Shaw:1993}. \citet{Mayer:2004} showed that bars need a minimum mass to form in galaxies with low surface brightness. Because the disk is diluted and not self-gravitating in a stabilizing halo, this leads to multi-arm or tightly wound spiral patterns with small $|\phi|$. \citet{Foyle:2008} demonstrated how the mass concentration can change the pattern and also create two exponential disks. 

A decade of research has demonstrated a strong correlation between pitch angle and black hole mass \citep{Seigar:2008,Ringermacher:2009,Berrier:2013,Davis:2017}. Combining the fact that the geometry of spiral arms is undeniably a consequence of galactic rotation \citep[and likely other factors such as the bulge mass and the density of the disk, see][]{Davis:2015} with the results of these aforementioned studies of $M_{\rm DM}$, $M_{\rm BH}$, $\phi$, and shear, inevitably lead to the conclusion that an $M_{\rm DM}$--$\phi$ relation must exist. Notably, the total gravitational mass of a galaxy determines its $v_{\rm rot}(R)$ profile, shear rate is a parameterization of the differential rotation exhibited by the $v_{\rm rot}(R)$ profile, which establishes a quasi-static density wave and generates a long-lived logarithmic spiral pattern of enhanced star formation with a unique pitch angle. 

Moving farther afield, high-$z$ spirals are rare but documented in the literature. \citet{Ivo:2003} presented a sample of half a dozen ``large disk-like galaxies at high redshift'' ($1.4 \leq z \leq 3.0$), with some of the nearer galaxies exhibiting ``evidence of well-developed grand-design spiral structure.'' \citet{Contini:2016} studied a sample of 28 star-forming galaxies at $0.2<z<1.4,$ some of which were identified to exhibit spiral structure. Similarly, \citet{Burkert:2016} analyzed a sample of 433 ($0.76<z<2.6$) massive star-forming disk galaxies. Recently, \citet{Yuan:2017} discovered a gravitationally lensed high-redshift spiral galaxy at $z=2.54$ through source-plane reconstruction, making it ``the highest-redshift spiral galaxy observed with the highest spatial resolution and spectroscopic depth to date.''\footnote{See also the potential evidence for spiral arms in high-redshift submillimeter galaxies from \citet{Hodge:2018}.} Just a few years ago, spiral galaxies were thought not to exist beyond $z\sim1.8$, and the spirals found at $z\gtrsim1$ were thought not to be \emph{true} spirals, merely ``spiral-like alignments of star formation clumps'' \citep{Elmegreen:2014}. Studies of these early epochs are essential for evolutionary studies. \citet{Dokkum:2013} showed that mass growth in the central regions (and the black holes) of present-day Milky Way-like galaxies occurred primarily between $z=2.5$ and $z=1$.

The discovery of high-$z$ spiral galaxies allows for the derivation of the high-$z$ $v_{\rm max}$--$\phi$ and high-$z$ $v_{\rm max}$--$\sigma_0$ relations, which can be compared with their local counterparts, checking for signs of evolution. There \emph{may} also be implications for the evolution of the black hole mass. Recent results from \citet{Drew:2018} reveal a star-bursting galaxy at $z=1.6$ with a flat outer rotation curve, which is indicative of a rotation-supported disk galaxy, rich in dark matter and similar to disk galaxies at low redshift. If this single finding is representative of other high-$z$ galaxies, our $M_{\rm DM}$--$\phi$ relation may become a useful tool in the estimation of dark matter at earlier epochs than our local sample. Because logarithmic spirals are scale-invariant curves and are thus unaltered by distance, $\phi$ might be an ideal messenger to convey an estimate of a host galaxy dark matter mass, especially at high redshifts, \emph{if} the $M_{\rm DM}$--$\phi$ relation does not evolve with $z$, and we can therefore use the local $M_{\rm DM}$--$\phi$ relation.

\section{Conclusions}\label{end}

The $v_{\rm max}$--$\sigma_0$ relation (\S\ref{sec:v-sigma}) provides insights into the apparent connection between the kinematics of a galaxy dark matter halo and its stellar bulge. We have shown that the $v_{\rm max}$--$\sigma_0$ relation, for $\sigma_0\gtrsim100\,{\rm km\,s^{-1}}$, is consistent with studies over nearly the past two decades. We find that the slope is inconsistent with a value of one at the $3.5\,\sigma$ level, revealing that $v_{\rm max}/\sigma_0$ is not a constant value. We introduce and highlight the implied $M_{\rm DM}$--$\sigma_0$ relation (Equation~(\ref{eqn:M_DM-sigma})) and its practical application for galaxies with $\sigma_0$ measurements of their bulges, yielding predictions of their dark matter halo masses.

We have presented observational evidence that a strong correlation exists between $M_{\rm BH}$ and $v_{\rm max}$ for spiral galaxies (Figure~\ref{fig:M_BH-log_v}). The $M_{\rm BH}$--$v_{\rm max}$ and $M_{\rm BH}$--$\sigma_0$ relations for spiral galaxies with $\sigma_0\gtrsim100\,{\rm km\,s^{-1}}$ have similar levels of scatter, with $\Delta_{\rm rms} = 0.58\,{\rm dex}$ and $0.63\,{\rm dex}$ in $\log M_{\rm BH}$, respectively. \emph{Alleged} pseudobulges dominate our sample of spiral galaxies (36 out of 42). Some may therefore prefer $v_{\rm max}$, over $\sigma_0$, as a predictor of black hole masses for spiral galaxies. Our $M_{\rm BH}$--$v_{\rm max}$ relation (Equation~(\ref{eqn:M_BH-log_v})) is steeper than previous studies, but we argue that past studies did not demonstrate consistency between complementary scaling relations, i.e., the slope we present for the $M_{\rm BH}$--$v_{\rm max}$ relation is supported by its agreement with the unification of the Tully--Fisher ($M_{\rm *,tot}$--$v_{\rm max}$) relation (Equation~(\ref{eqn:TF})) and the $M_{\rm BH}$--$M_{\rm *,tot}$ relation (Equation~(\ref{eqn:paper2})).

The scatter for the $M_{\rm BH}$--$M_{\rm *,tot}$ relation \citep{Davis:2018} is higher ($\Delta_{\rm rms} = 0.79\,{\rm dex}$ in $\log M_{\rm BH}$), suggesting that the influence of dark matter, as traced by $v_{\rm max}$,\footnote{All relations in this work involving $M_{\rm DM}$ are constrained by the \citet{DC14} halo profile and the empirical $M_{\rm DM}$--$v_{\rm max}$ relation (Equation~(\ref{eqn:conversion})) of \citet{Katz:2018}. Other relations using $v_{\rm max}$ are based directly on observational data.} (rather than total stellar mass) on the central black hole mass is relevant. The low $\Delta_{\rm rms}$ value of just $0.43$\,dex in $\log M_{\rm BH}$ \citep{Davis:2017} about the shallow slope of the $M_{\rm BH}$--$\phi$ relation still remains unmatched for spiral galaxies via $\phi$, which is a defining characteristic of spiral galaxies. We acknowledge that these studies \citep[i.e.,][and this work]{Davis:2017,Davis:2018,Davis:2019} focus on a sample constructed from selection of all spiral galaxies with available dynamical mass estimates of their central black holes. Because the sample size (48) is not yet large enough to explore a volume-limited sample, the resulting scaling relations unavoidably suffer from selection effects. Thus, underrepresented populations of galaxies with difficult to measure black hole masses would likely act to increase the scatter in these relations if they could be included (i.e., the quoted scatters should be considered minimum estimates),

Because $v_{\rm max}$ is a reliable predictor of $M_{\rm DM}$ for late-type galaxies \citep{Katz:2018}, our strong $M_{\rm BH}$--$v_{\rm max}$ correlation ($r = 0.74$ and $r_s=0.72$) retains its strength when translated into an $M_{\rm BH}$--$M_{\rm DM}$ relation (Equation~(\ref{eqn:M_BH-M_DM})). We find steep slopes for these two relations, which is indicative of a history of substantial black hole mass increase through accretion and long-lived quasars \citep{Haehnelt:1998}. Concerning late-type galaxies, evolution along the steep $M_{\rm BH}$--$M_{\rm *,tot}^{3.05\pm0.53}$ relation implies that galaxies host black holes that underwent rapid mass growth relative to their host (i.e., relative to maintaining a constant $M_{\rm BH}/M_{\rm *,tot}$ ratio). We echo our arguments from \citet{Davis:2018} that it is crucial to separate galaxies according to morphology when black hole mass scaling relations are investigated. As revealed in \citet{Savorgnan:2016:II} and \citet{Sahu:2019}, the $M_{\rm BH}$--$M_{\rm *,tot}$ relation for early-type galaxies is very different.

We have also provided evidence of a link between the geometry of spiral galaxy arms and their dark matter halos. Indeed, the seminal work by \citet{Kennicutt:1981} qualitatively showed that $v_{\rm max}$ is well correlated with $\phi$, in particular for the later-type spiral galaxies. Such a specific connection between the kinematics of a galaxy and its spiral pattern has important consequences for spiral-arm genesis theories, such as the spiral density-wave (quasi-stationary spiral structure) modal theory \citep{Lindblad:1963,Lindblad:1964,Lin:Shu:1964,Lin:Shu:1966,Roberts:1975,Roberts:1975b,Bertin:1989,Bertin:1989b,Bertin:1991,Bertin:1993,Bertin:1996,Fuchs:1991,Fuchs:2000,Bertin:Lin:1996}, swing amplification theory \citep{Julian:1966,Goldreich:1978,Toomre:1981,Toomre:1991,D'Onghia:2013}, kinematic spiral waves \citep{Kalnajs:1973,Toomre:1977}, the stochastic self-propagating star formation theory \citep{Gerola:1978,Seiden:1979}, solid-body rotation of material arms \citep{Kormendy:1979}, recurrent cycles of groove modes \citep{Sellwood:2000}, manifold theory \citep[][see also \citealt{Danby:1965}]{Romero:2006,Voglis:2006}\footnote{Two versions of the manifold theory have been developed: (i) the ``flux-tube'' version \citep{Romero:2006,Romero:2007,Athanassoula:2009b,Athanassoula:2009,Athanassoula:2010,Athanassoula:2012} and (ii) the ``apocentric manifolds'' version \citep{Voglis:2006b,Voglis:2006,Tsoutsis:2008,Tsoutsis:2009,Efthymiopoulos:2010,Harsoula:2016}; see also the recent work by \citet{Efthymiopoulos:2019}.}, superposed transient instabilities \citep{Sellwood:2014}, etc. We have quantified the $v_{\rm max}$--$\phi$ relation (Equation~(\ref{eqn:phi-log_v_rot})) to enable and facilitate testing between rival theories of spiral-arm formation \citep{Kennicutt:1982}. The connection between spiral-arm winding and kinematics can also be applied to the Tully--Fisher relation to indirectly estimate distances to spiral galaxies (Equation~(\ref{eqn:phi-mag})). For a complete understanding of galaxy formation and evolutionary processes, it is necessary to study and gain insight into the influence of dark matter. To that end, we have translated the $v_{\rm max}$--$\phi$ relation into an $M_{\rm DM}$--$\phi$ relation (Equation~(\ref{eqn:M_DM-phi})) providing for the first time a useful method of predicting dark matter halo masses from the geometry of spiral arms in uncalibrated photometric images of spiral galaxies.

Metaphorically, we can ascribe a dark matter halo as the central nervous system of its host galaxy, filling and connecting all regions of the galaxy. Indeed, the dark matter halo of a spiral galaxy can be considered a legitimate puppet-master, covertly influencing the observable properties and components of its host galaxy. This notion that the dark matter halo determines everything in a spiral galaxy (e.g., $v_{\rm max}$, mass concentration, and maybe $M_{\rm BH}$) has already been put forward by the proponents of modified gravity \citep{Milgrom:1983,McGaugh:2000}: when the baryon distribution is known, the implied dark matter distribution and total mass can be deduced.\footnote{From the standpoint of modified Newtonian dynamics, \citet{Milgrom:1983} would argue that the baryonic mass \emph{is} the total mass (i.e., there is no dark matter).} Much of this dates back to the Tully--Fisher relation \citep{Tully:Fisher:1977,Aaronson:1979,Aaronson:1980,Mould:1980,Aaronson:1983}, refined by the baryonic Tully--Fisher relation \citep{Freeman:1999,Walker:1999,McGaugh:2000}.\footnote{This critical dependency on the inventory of baryons would also necessitate an accounting of the gas in galaxies as well (i.e., taking gas into account should improve all scaling relations). For our sample, the gas content is thought to be a small fraction of the total mass, and thus we ignore it.}

Intermediate-mass black holes (IMBHs; $10^2\,M_\sun\leq M_{\rm BH}<10^5\,M_\sun$) remain elusive stepping stones in the evolution from stellar mass ($M_{\rm BH}<10^2\,M_\sun$) to supermassive ($M_{\rm BH}\geq10^5\,M_\sun$) black holes (see \citealt{Mezcua:2017} and \citealt{Koliopanos:2017b} for reviews). Recent studies \citep{Koliopanos:2017,Graham:2019,Graham:2019b} have used multiple black hole mass scaling relations, in combination with X-ray observations of nuclear point sources, to identify IMBH candidates in local galaxies. Studies like these provide several independent estimates of black hole masses and serve as important guides for follow-up targets to hopefully confirm IMBH masses through more direct measurements. We offer one additional black hole hole mass scaling relation: $M_{\rm BH}$--$v_{\rm max}$, which, when extrapolated, predicts that galaxies with $v_{\rm max}<130\,{\rm km\,s^{-1}}$ ($\log\left[M_{\rm DM}/M_\sun\right]<11.49$ through Equation~(\ref{eqn:M_BH-M_DM})) should harbor IMBHs. For reference, our Galaxy\footnote{The quantity $v_{\rm max}$ should not be confused with the circular rotation velocity of the disk of our Galaxy at the location of the Sun and equal to $(238\pm15)\,{\rm km\,s^{-1}}$ \citep{Bland-Hawthorn:2016} or $(229.0\pm0.2)\,{\rm km\,s^{-1}}$ \citep{Eilers:2019}.} has $v_{\rm max}=\left(198\pm6\right)\,{\rm km\,s^{-1}}$ and $\log\left(M_{\rm DM}/M_\sun\right)=11.86\pm0.01$ \citep{Eilers:2019}, with $\log\left(M_{\rm BH}/M_\sun\right)=6.60\pm0.02$ \citep{Boehle:2016}.

Despite claims that $M_{\rm BH}$ does not correlate with disks \citep{Kormendy:2011}, the $M_{\rm BH}$--$v_{\rm max}$ relation for spiral galaxies has similar scatter as the $M_{\rm BH}$--$\sigma_0$ relation for spiral galaxies with $\sigma_0\gtrsim100\,{\rm km\,s^{-1}}$. The steep $M_{\rm BH}$--$v_{\rm max}$ relation observed here brings consistency between the Tully--Fisher relation ($M_{\rm *,tot}\propto v_{\rm rot}^{4.0\pm0.1}$) and the $M_{\rm BH}\propto M_{\rm *,tot}^{3.05\pm0.53}$ relation for spiral galaxies. We note that the  $M_{\rm BH}$--$v_{\rm max}$ relation (Equation~(\ref{eqn:M_BH-log_v})) and the $v_{\rm max}$--$\sigma_0$ relation (Equation~(\ref{eqn:log_v-log_sigma})) are also compatible with the $M_{\rm BH}$--$\sigma_0$ relation for spiral galaxies \citep{Davis:2017}, i.e., $M_{\rm BH}\propto v_{\rm max}^{10.62\pm1.37}$ combined with $v_{\rm max}$--$\sigma_0^{0.65\pm0.10}$ yields $M_{\rm BH}\propto \sigma_0^{6.87\pm1.42}$, which is consistent with our result for spiral galaxies ($M_{\rm BH}\propto \sigma_0^{5.65\pm0.79}$) from \citet{Davis:2017} at the level of $0.55\,\sigma$. Alternatively, if we cautiously assume that a selection bias is present, censoring galaxies with low $\sigma_0$ from our sample, we can instead combine Equation~(\ref{eqn:M_BH-log_v}) with Equation~(\ref{eqn:log_v-log_sigma_alt}), which yields $M_{\rm BH}\propto \sigma_0^{3.21\pm1.19}$. When we compare this with the (Y$|$X) regression fit to the $M_{\rm BH}$--$\sigma_0$ relation (which similarly reduces any potential bias due to unobserved galaxies with low velocity dispersions) from \citet{Davis:2017}, $M_{\rm BH}\propto \sigma_0^{3.88\pm0.89}$, there is consistency at the level of $0.32\,\sigma$. To conclude, the self-consistent relations presented in this work were derived from a spiral galaxy sample with directly measured SMBH masses that doubles the size of previous samples. These relations necessitate a salient revision to constraints that are applied to theory and simulations.

\acknowledgments

We thank Duncan Forbes and Victor Debattista for feedback that helped improve the presentation of this paper. The Australian Research Council's funding scheme DP17012923 supported A.W.G. Parts of this research were conducted by the Australian Research Council Centre of Excellence for Gravitational Wave Discovery (OzGrav), through project number CE170100004. This research has made use of NASA's Astrophysics Data System. We acknowledge use of the HyperLeda database (\url{http://leda.univ-lyon1.fr}).

\bibliography{bibliography}

\end{document}